\documentclass[fleqn,usenatbib]{mnras}
\usepackage{newtxtext,newtxmath}
\usepackage[T1]{fontenc}

\usepackage{graphicx}	
\usepackage{amsmath}	
\DeclareMathOperator\erf{erf}
\title[Dwarf stellar haloes]{Dwarf stellar haloes: a powerful probe of small-scale galaxy formation and the nature of dark matter}

\author[Alis J. Deason et al.]{
Alis J. Deason,$^{1,2}$\thanks{E-mail: alis.j.deason@durham.ac.uk}
Sownak Bose,$^{1,3}$
Azadeh Fattahi,$^{1}$
Nicola C. Amorisco,$^{1}$
Wojciech Hellwing,$^{4}$\newauthor
and Carlos S. Frenk$^{1}$
\\
$^{1}$Institute for Computational Cosmology, Department of Physics, Durham University, Durham DH1 3LE, U.K\\
$^{2}$Centre for Extragalactic Astronomy, Department of Physics, University of Durham, South Road, Durham DH1 3LE, UK\\
$^{3}$Center for Astrophysics, Harvard \& Smithsonian 60 Garden St. Cambridge, MA 02138, USA\\
$^{4}$Center for Theoretical Physics, Polish Academy of Sciences, Al. Lotników 32/46, 02-668 Warsaw, Poland
}

\date{Accepted XXX. Received YYY; in original form ZZZ}

\pubyear{2021}

\begin{document}
\label{firstpage}
\pagerange{\pageref{firstpage}--\pageref{lastpage}}
\maketitle

\begin{abstract}
We use $N$-body cosmological simulations and empirical galaxy models to study the merger history of dwarf-mass galaxies (with $M_{\rm halo}\sim10^{10}M_\odot$). Our input galaxy models describe the stellar mass-halo mass relation, and the galaxy occupation fraction. The number of major and minor mergers depends on the type of dark matter; in particular, minor mergers are greatly suppressed in warm dark matter models. In addition, the number of mergers that bring in stars is strongly dependent on the galaxy occupation model. For example, minor mergers are negligible for stellar halo growth in models with a high mass threshold for galaxy formation (i.e. $10^{9.3}M_\odot$ at $z=0$). Moreover, this threshold for galaxy formation can also determine the relative difference (if any) between the stellar haloes of satellite and field dwarfs. Using isolated simulations of dwarf-dwarf mergers, we show that the relative frequency of major and minor mergers predict very different stellar haloes:  Typically, ``intermediate'' dark matter merger ratios ($\sim$ 1:5) maximize the growth of distant stellar haloes. We discuss the observability of dwarf stellar haloes and find that the surface brightness of these features are incredibly faint. However, when several dwarfs are stacked together, models that form particularly rich stellar haloes could be detectable. Finally, we show that stellar streams in the Galactic halo overlapping in phase space with known dwarf satellites are likely remnants of their stripped stellar haloes. The mere existence of dwarf stellar haloes can already put constraints on some small-scale models, and thus observational probes should be a high priority.

\end{abstract}


\section{Introduction}
The hierarchical nature of structure formation predicts that haloes grow from the aggregation of several lower mass ``clumps'' \citep{frenk88}. A natural consequence of this process is the existence of an extended halo of stars surrounding galaxies, which is built up from the debris of destroyed lower mass objects. The nature of these haloes has been studied extensively from Milky Way-mass to cluster-mass systems \citep[e.g][]{bullock05,mihos05,abadi06, conroy07, purcell07,cooper10}. However, the stellar haloes of dwarf galaxies have been given relatively little attention. Indeed, it remains an open question as to whether or not stellar haloes around dwarf galaxies even exist.

The best-studied stellar halo is probably our own Galactic one. Here, we have access to a star-by-star view of this meagre stellar component. The Galactic stellar halo only comprises $\sim 1\%$ of the stellar mass of the Milky Way \citep[e.g][]{deason19, mackereth20}, but, arguably, provides the most information on the Galactic assembly history and the underlying dark matter potential \citep[see e.g.][for recent reviews]{bland-hawthorn16, helmi20}. The combination of hierarchical assembly and the (relatively) steep stellar mass-halo mass relation below $10^{12}M_\odot$ naturally leads to (accreted) Galactic stellar haloes having only a few main progenitors \citep[e.g][]{cooper10, deason16, amorisco17b, fattahi20b}. This is also borne out observationally, with recent evidence suggesting that the inner halo is dominated by one massive merger event \citep{belokurov18, helmi18}. However, there is a strong radial dependence, and it is predicted that the outer reaches of the halo are likely populated by several, lower mass progenitors \citep[e.g.][]{monachesi19, fattahi20b}.

On dwarf mass scales, as mentioned above, the stellar haloes are much less well studied. This is for several reasons. First, the fraction of mass assembled via mergers reduces at low mass scales, and is instead dominated by ``smooth'' accretion \citep[e.g.][]{genel10}. Second, while for Milky Way ($10^{12}M_\odot$) and perhaps Magellanic Cloud ($10^{10-11}M_\odot$) mass-scales the stellar mass-halo mass (SMHM) relation is reasonably well constrained \citep[e.g.][]{behroozi13, moster13}, there is much greater uncertainty at lower masses. For example, assuming a monotonic relation $M_{\rm star} \propto M^\alpha_{\rm halo}$, slopes between $\alpha=1.4$ and $3.1$ have been quoted in the literature for low halo masses $M_{\rm halo} \lesssim 10^{10}M_\odot$ \citep[e.g.][]{brook14, garrison-kimmel14, read17}. Moreover, the scatter at the low-mass end is highly uncertain. At higher masses ($> 10^{10}M_\odot$), a 0.2-0.3 dex scatter in $\mathrm{log} M_{\rm star}$ at fixed halo mass is generally accepted \citep[e.g.][]{behroozi13,moster13}. However, several works have suggested that the scatter is much larger at low mass scales and may even increase as halo mass decreases \citep[e.g.][]{garrison-kimmel17, munshi21}. Third, and to complicate matters, it is not only the mass of stars in a given halo that matters - it is also a question of whether the stellar component exists at all! Not all dark matter haloes host galaxies, and this occupation fraction strongly depends on halo mass. However, the mass scale at which dark haloes are ``starless'' is under serious debate, with halo mass thresholds at $z=0$ ranging from $10^{7.5}$ to $10^{9.3}M_\odot$ quoted in the literature \citep[e.g.][]{sawala16, fitts18, jethwa18, graus19, wheeler19, benitez-llambay20, nadler20, kravtsov21}.

The physical interpretation of the dwarf SMHM relation, and the galaxy occupation depend on two fundamental influences on small-scale galaxy formation: the star formation and stellar feedback at low mass scales, and the effect of reionization. It has been realised for some time that the epoch of reionization is a vital influence on the number counts of surviving dwarf galaxies. Indeed, the apparent mismatch between counts of observable satellite dwarf galaxies in the Milky Way and subhaloes in numerical simulations (i.e. the ``missing satellites problem''), is largely due to the influence of reionization raising the gas pressure and temperature, and thus suppressing star formation at low mass scales \citep[e.g.][]{bullock00, benson03}. More recently, \cite{bose18} showed that reionization is imprinted in the luminosity function of dwarf galaxies, and that the epoch of reionization, and the dwarf mass-scale that is influenced by this event, can potentially be derived from the number counts of dwarf galaxies.

The nature of the dark matter particle also has a significant impact on dwarf mass scales. For example, warm dark matter (WDM) particles have non-negligible thermal velocities at early times, which leads to a strong suppression of the linear matter power spectrum on low mass scales \citep[e.g.][]{dodelson94, bode01, zentner03, benson13}.
Cosmological $N$-body WDM simulations show that the number counts of haloes differ substantially from CDM for masses below $\sim 10^9M_\odot$ \citep[e.g.][]{colin00, schneider12, lovell14, bose16, wang17, lovell21}. Indeed, invoking WDM rather than cold dark matter (CDM) is an alternative solution to the missing satellites problem, without relying on baryonic influences. Of course, when probing the number counts of dwarf galaxies there is an inherent degeneracy between the dark matter and baryonic models, and a robust answer will require several independent lines of evidence.

While the observed dwarf satellite galaxies present an excellent point of comparison for fundamental dark matter and galaxy formation theories, the affect of halo-to-halo scatter, and environmental effects mean that several more ``data points'' are very much sought-after. Clearly, an obvious, but largely unexplored, avenue is to consider destroyed dwarf galaxies - particularly, those related to known surviving dwarfs. Recent work by \cite{chiti21} has been particularly persuasive at bringing dwarf stellar haloes to the forefront. These authors find that the Tucana II ultra-faint dwarf galaxy has an unusually extended stellar distribution, that could potentially be related to (early) merger events. Indeed, \cite{tarumi21} showed that early major mergers can form an extended stellar halo similar to what is seen in Tucana II. On slightly more massive scales, \cite{johnson20} use the H3 survey \citep{conroy19} to find a kinematically diffuse, low metallicity population related to the disrupting Sagittarius (Sgr) dwarf galaxy. These authors suggest that this could be the stellar halo of the Sgr dwarf progenitor. These compelling results, in addition to several other lines of evidence for ``stellar halo-like'' populations of dwarf galaxies \citep[e.g.][]{coleman04, martinez-delgado12, pace20, qi21}, suggest that this avenue for exploring the lowest mass galaxies is starting to bear fruit.

Previous theoretical work exploring the accreted stellar populations of dwarf galaxies have focused on a small number of dwarf-mass haloes, and/or considered only one galaxy formation model \citep[e.g.][]{fitts18, kado-fong21, martin21, tarumi21}. In this work we take a different approach. Namely, we use a large suite of high resolution, dark matter only simulations and consider a wide range in galaxy formation models. This ensures we can \textit{statistically} examine the stellar accretion events of dwarf galaxies, and discern how different galaxy models affect the resulting dwarf stellar haloes. Moreover, we also explore both cold \textit{and} warm dark matter models; at the mass scales we are investigating the nature of the dark matter particle can have a significant influence \citep[e.g.][]{lovell14,bose16}. In Section \ref{sec:coco} we describe the cosmological $N$ body simulations used in this work, and follow the merger histories of $10^{10}M_\odot$ mass haloes at $z=0$. Empirical galaxies models, which include the SMHM relation, and the galaxy occupation model, are introduced in Section \ref{sec:gals}. We use these empirical models, in combination with the cosmological $N$ body simulations to probe the accreted \textit{stellar} mass of dwarf galaxies. In Section \ref{sec:gadget} we use isolated $N$ body simulations to investigate how different merger events can build up a dwarf stellar halo. The observability of such stellar haloes is discussed in Section \ref{sec:obs}, and finally, we summarise our main conclusions in Section \ref{sec:conc}.

\section{Dark Matter Simulations}
\label{sec:coco}
In this section, we explore the merger rates of low-mass ($\sim 10^{10} M_{\odot}$) dark matter subhaloes using cosmological $N$-body simulations. The advantage of using dark matter only simulations is that we can probe large volumes with high resolution, which is essential in order to probe down to dwarf galaxy scales. We discuss the link between these dark matter subhaloes and dwarf galaxies in Section \ref{sec:gals}.
\begin{figure*}
	\includegraphics[width=\linewidth]{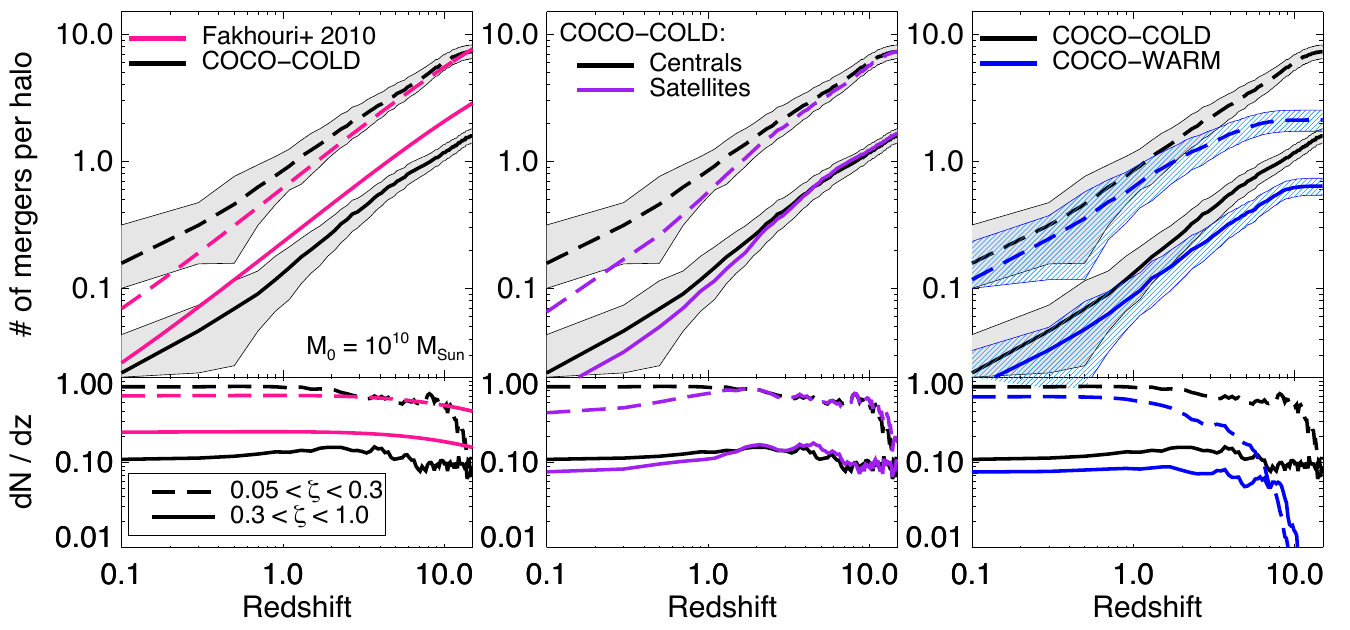}
    \caption[]{The number of mergers (per halo) as a function of redshift for $10^{10}M_\odot$ mass haloes in the \textsc{coco} simulations. The top panels show the cumulative number, and the bottom panels show the differential number ($\mathrm{d}N/\mathrm{d}z$). The solid lines indicate major mergers with mass ratio $0.3 < \zeta < 1$, and the dashed lines show minor mergers with mass ratio $0.05 < \zeta < 0.3$. The shaded regions for the cumulative numbers indicate the $1\sigma$ uncertainties (assuming Poisson noise). In the \textit{left-hand panel} we show the analytical (CDM) prediction from \cite{fakhouri10} with the pink coloured lines. There is good agreement for minor mergers, but \textsc{coco} predicts slightly fewer major mergers at this mass scale. In the \textit{middle-panel} we show the cumulative number of mergers for satellite systems in \textsc{coco-cold} with the purple lines (centrals are shown with the black lines, as in the left and right-hand panels). The total number of mergers is very similar, but it is unusual for satellite systems to undergo mergers at low redshift (typically, after infall onto a host halo). The \textit{right-hand panel} compares the CDM and WDM models. The number of both major and minor mergers is smaller in \textsc{coco-warm} relative to \textsc{coco-cold}.}
    \label{fig:dm_mergers}
\end{figure*}

\subsection{Copernicus Complexio (\textsc{coco})}
The $N$ body simulations used in this work are the Copernicus Complexio (\textsc{coco}) suite, which were first introduced in \cite{bose16} and \cite{hellwing16}. \textsc{coco} is a zoom-in re-simulation of a patch of a (100 Mpc)$^3$ low resolution volume (Copernicus complexio Low Resolution, \textsc{color}). The high resolution volume is approximately (40 Mpc)$^3$, with dark matter particle mass, $m_p = 1.6 \times 10^5M_\odot$, and force softening $\epsilon = 0.33$ kpc. \textsc{coco} was run using the \textsc{Gadget-3} code \citep{springel01b,springel05} from $z=127$ to $z=0$, and dark matter particle velocities and positions were saved for 160 equally spaced snapshots in $\mathrm{log}(1+z)$. \textsc{coco} assumes cosmological parameters derived from WMAP-7 \citep{komatsu11}, with $\Omega_{\rm m} =0.272$, $\Omega_{\Lambda}=0.728$, $h=0.704$, $n_s=0.967$ and $\sigma_8=0.81$. Note that \textsc{coco} consists of two sets of simulations: one with a CDM particle, and another with a WDM particle. The WDM model (with rest mass 3.3 keV) is compatible with current Lyman $\alpha$ constraints \citep{viel13}, but is the `warmest' particle allowed by these measures. It is also very similar to the coldest sterile neutrino model compatible with a particle \citep{lovell16}. In this work, we make use of both simulations and refer to them as \textsc{coco-cold} and \textsc{coco-warm}, respectively. 

Dark matter haloes and subhaloes were identified using the \textsc{subfind} algorithm \citep{springel01a}. Haloes are initially identified using the friends-of-friends (FOF) algorithm \citep{davis85}, and \textsc{subfind} then analyses each FOF group to find self-bound dark matter subhaloes. Throughout this work, the subhalo mass is defined as the mass identified by \textsc{subfind} that is gravitationally bound to the subhalo.

\subsection{Dark matter mergers}

We now explore the merger histories of $10^{10}M_\odot$ mass haloes in the \textsc{coco} simulations (which are resolved with $\sim 10^5$ particles). Here, we consider haloes with \textit{peak} subhalo mass in the range $10^{9.8} < M^{\rm peak}_{\rm halo}/M_\odot < 10^{10.2}$, where ``peak'' mass is defined as the maximum mass that a subhalo reaches over cosmic time. It has been shown that the peak mass is a much better proxy for stellar mass than $z=0$ subhalo mass \citep[e.g.][]{conroy06, behroozi13, moster13}. Moreover, this definition allows us to compare central and satellite subhaloes at $z=0$ of similar stellar mass (and hence similar $M_{\rm peak}$). It is particularly important to consider $M_{\rm peak}$ for satellite galaxies, as they can undergo significant tidal stripping, and their $z=0$ halo masses can be much lower than their peak mass.
By default, we consider central haloes at $z=0$, and there are $N=3981 (3822)$ haloes in \textsc{coco-cold} (\textsc{coco-warm}) that satisfy our selection criteria. For comparison, we also explore the $z=0$ satellite population within the same $M_{\rm peak}$ range. There are $N=1933 (1971)$ equivalent satellite systems in \textsc{coco-cold} (\textsc{coco-warm}). Note that we do not consider the central and satellite systems together as they are fundamentally two different populations that undergo different merger histories (despite having similar $M_{\rm peak}$ values). This is explored in more detail in Section \ref{sec:sats}. Unless otherwise stated, we show the properties of central haloes at $z=0$ throughout this work.

For each of the $10^{10}M_\odot$ mass haloes we track them back in time and identify all the mergers that occur. Here, we consider a merger to happen when a subhalo belonging to the main progenitor no longer appears in the halo catalogue, and the particles belonging to that descendant subhalo become part of the main progenitor. Note that the definition of \textit{when} a merger occurs is fairly arbitrary, but we assume that the time at which the dark matter halo falls below the resolution limit in \textsc{subfind} is a reasonable approximation for the time at which the \textit{stellar} components of these subhaloes coalesce (cf. \citealt{deason14}).

The $z=0$ halo mass function of \textsc{coco} shows good agreement with analytical fits to the CDM mass function \citep{sheth99, jenkins01} down to halo masses of $\sim 5 \times 10^{6}M_\odot$ (see Fig. 7 in \citealt{bose16}). The mass resolution of \textsc{coco} allows us to track $10^{10}M_\odot$ mass subhaloes at $z=0$ back to high redshift, and explore both their major and minor mergers. For example, a typical halo in our sample has a mass of $\approx 10^8M_\odot$ at $z\sim 10$. Thus, we can study minor mergers with mass-ratio 1:20 up to this redshift (i.e. at $z \sim 10$ a 1:20 merger event is resolved with $\sim \! 600$ and $\sim \! 30$ particles, respectively). However, the numbers of minor mergers at very high redshift ($z \gtrsim 10$) are likely lower limits. Indeed, the rapid drop-off in minor merger rates at $z \gtrsim 10$ (see bottom panels of Fig. \ref{fig:dm_mergers}) is likely a consequence of resolution effects.  Encouragingly, our estimated numbers of minor mergers for $z \lesssim 10$ are in good agreement with analytical predictions (see text below, and Fig. \ref{fig:dm_mergers}), so we do not believe our results are strongly influenced by numerical effects in this regime.   

The mass-ratios of the mergers, $\zeta = M^{\rm Satellite}_{\rm halo}/ M^{\rm Central}_{\rm halo}$,  are defined by considering the peak mass of the merging satellite, and the mass of the central at the time of the merger. Again, we use $M_{\rm peak}$ here as the satellites have normally undergone significant stripping before being destroyed, and their masses just before destruction are significantly lower than their peak mass. Throughout this work, the mass-ratios of the mergers, $\zeta$, are defined based on the dark matter masses of the merger events. In Fig. \ref{fig:dm_mergers} we show the number of mergers (per halo) as a function of redshift for the $10^{10}M_\odot$ mass haloes in our sample. The top panels show the cumulative number, and the bottom panels show the differential number ($\mathrm{d}N/\mathrm{d}z$) as a function of redshift. The solid lines indicate major mergers with, $0.3 < \zeta < 1$, and the dashed line indicates minor mergers with $0.05 < \zeta < 0.3$. For comparison, in the left-panel of Fig. \ref{fig:dm_mergers} we show the analytical estimates of these (CDM) merger rates from \cite{fakhouri10} derived from the Millennium-II simulation. There is good agreement for minor mergers, but the number of major mergers is slightly lower in \textsc{coco} (by $\sim 30\%$). There are a number of possible reasons for the (relatively) low major merger rates. \textsc{coco} is a higher resolution simulation with a different cosmology to Millennium-II, and we are also adopting a different merger mass-ratio definition (i.e. peak mass) in our approach. In addition, the \textsc{coco} zoom-in region is slightly underdense compared to the cosmic mean, and there is a clear environmental trend with the formation redshift of haloes \citep[e.g][]{hellwing21}. Indeed, \cite{hopkins10} showed that subtle changes in mass definition, environment, cosmology etc. can lead to variations in halo-halo merger rates by factors of $\sim2$. 

The middle panel of Fig. \ref{fig:dm_mergers} compares the merger properties of central and satellite systems at $z=0$ in \textsc{coco-cold}. The overall number of mergers are very similar, but \textit{when} these mergers occur can differ. In particular, both major and minor mergers are less common in satellite systems at low redshift, but they can be slightly more common at higher redshift. This is because mergers are unlikely for satellite systems after infall onto a host halo. Moreover, the frequency of mergers can be higher for $z=0$ satellite dwarfs at higher redshift because the satellites reach their peak mass ($M_{\rm peak}$) at earlier times. Conversely, central haloes can accrete mass at later times and typically reach their peak mass at $z \approx 0$. This subtle difference between centrals and satellites, despite having similar $M_{\rm peak}$ or $M_{\rm star}$ at $z=0$, can have important implications for their accreted \textit{stellar} properties. We explore this more thoroughly in Section \ref{sec:sats}.

Our estimates from the \textsc{coco-cold} simulation suggest that $10^{10}M_\odot$ haloes at $z=0$ have typically undergone $N \sim 1.6$ major mergers, and $N \sim 7$ minor mergers. The merger frequencies are reduced by a factor of $\sim 3$ in \textsc{coco-warm}, with the dwarf-mass haloes experiencing $N \sim 0.6$ major mergers, and $N \sim 2$ minor mergers (see the right-hand panel of Fig. \ref{fig:dm_mergers}). In the following section, we introduce empirical galaxy models that allows us to link these dark matter merger rates to the $\textit{stellar}$ merger rates of these dwarf galaxies.

\section{Empirical galaxy models}
\label{sec:gals}

\begin{figure*}
	\includegraphics[width=\linewidth]{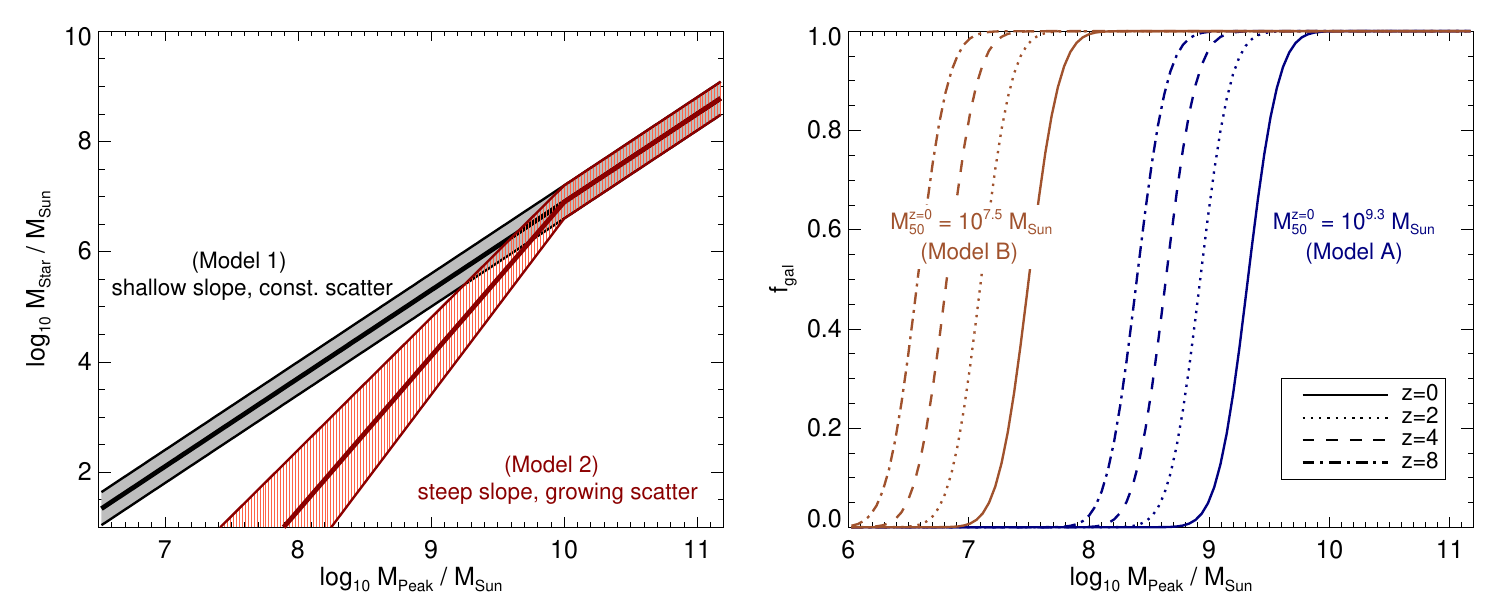}
    \caption{The empirical galaxy models used in this work. The left-hand panel shows the stellar mass-halo mass relation, and the right-hand panel shows the (redshift-dependent) galaxy occupation fraction. The models are chosen to roughly cover the range of models adopted in the literature. The parameters used in these models are given in Table \ref{tab:models}.}
    \label{fig:gal_models}
\end{figure*}

\begin{table*}
\begin{center}
\renewcommand{\tabcolsep}{0.4cm}
\renewcommand{\arraystretch}{1.8}
\begin{tabular}{| l | c | c| r |}
 \hline 
 \textbf{Model} & \textbf{SMHM} & \textbf{Galaxy occupation} \\
 \hline
 \hline
Model A1 & $\alpha=1.6$, $\sigma=0.3$ & $M^{z=0}_{50}=10^{9.3}M_\odot$, $c_1=-0.226, c_2=0.014$, $\sigma_{\rm gal}=0.2$ \\
Model A2 & $\alpha_1=1.6$, $\alpha_2=2.8$, $\sigma_0=0.3$, $\gamma=-0.4$ & $M^{z=0}_{50}=10^{9.3}M_\odot$, $c_1=-0.226, c_2=0.014$, $\sigma_{\rm gal}=0.2$\\
Model B2 & $\alpha_1=1.6$, $\alpha_2=2.8$, $\sigma_0=0.3$, $\gamma=-0.4$& $M^{z=0}_{50}=10^{7.5}M_\odot$, $c_1=-0.226, c_2=0.014$, $\sigma_{\rm gal}=0.2$ \\
Model B1 & $\alpha=1.6$, $\sigma=0.3$ & $M^{z=0}_{50}=10^{7.5}M_\odot$, $c_1=-0.226, c_2=0.014$, $\sigma_{\rm gal}=0.2$\\
\hline
  \end{tabular}
  \caption{Parameters for the four different empirical galaxy models used in this work.}
\label{tab:models}
\end{center}
\end{table*}

In this Section, we transform the halo mergers derived from \textsc{coco} to the predicted galaxy mergers. For this, we use empirical galaxy models that can be straightforwardly applied to the dark matter merger trees. Note, here we only consider stellar mergers (i.e. dry mergers) and do not consider the contribution of gas to the merger events. Our galaxy models comprise two components: (1) a stellar mass-halo mass (SMHM) relation, where stellar mass can be mapped on to (peak) halo mass analytically and (2) a galaxy occupation model which describes the probability of a dark matter halo hosting a galaxy. 

\begin{figure*}
  \begin{minipage}{0.49\linewidth}
        \centering
        \includegraphics[width=\textwidth,angle=0]{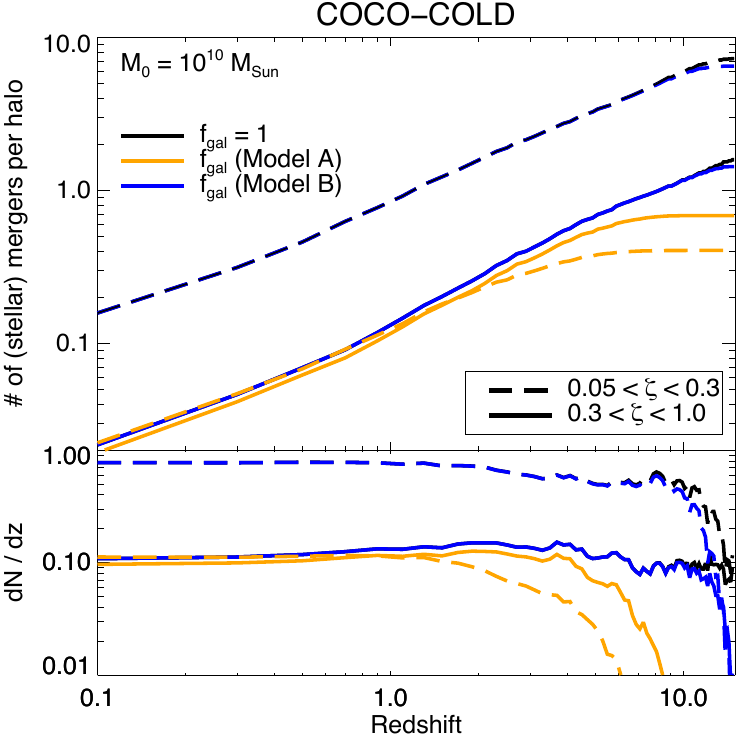}
    \end{minipage}
    \begin{minipage}{0.49\linewidth}
        \centering
        \includegraphics[width=\textwidth,angle=0]{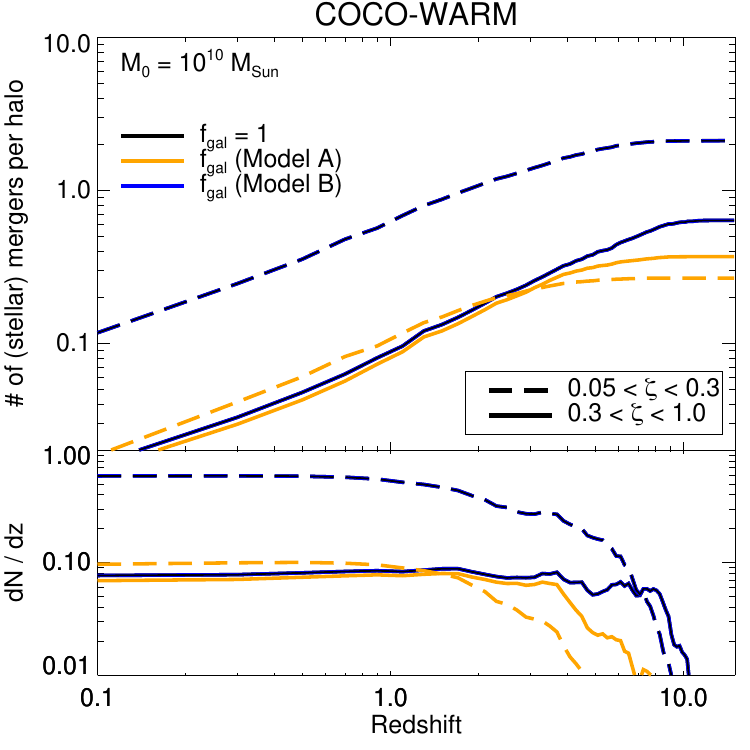}
    \end{minipage}
    \caption{The number of \textit{stellar} mergers (per halo) as a function of redshift for \textsc{coco-cold} (left-hand panel) and \textsc{coco-warm} (right-hand panel). As in Fig. \ref{fig:dm_mergers}, the solid/dashed lines indicate major/minor mergers, where $\zeta$ denotes the (dark matter) mass ratio. Here, we show different galaxy occupation fractions with coloured lines. The black lines with $f_{\rm gal} =1$ are the same as in Fig. \ref{fig:dm_mergers}. Models with a high mass threshold for galaxy formation (e.g. Model A) have a significantly reduced number of stellar mergers; this is particularly true for minor mergers.}
    \label{fig:dm_mergers_fgal}
\end{figure*}

We use two different SMHM relations (labelled with ``1'' and ``2''). Both models are normalized to have the same stellar mass ($10^{6.9}M_\odot$) at $M_{\rm halo} = 10^{10}M_\odot$. This stellar mass ($\sim 10^7 M_\odot$) roughly corresponds to the Fornax dwarf satellite of the Milky Way \citep{mcconnachie12}. This normalization of the SMHM relation coincides with the \cite{moster13} abundance matching relation, and also agrees with constraints from hydrodynamical simulations \citep[e.g.][]{munshi13, sawala15}.  Moreover, we find that in \textsc{coco}, Milky Way-mass systems (with $M_{\rm halo} \sim 0.9-1.6 \times 10^{12}M_\odot$) typically have $2 \pm 2$ satellite systems with $M_{\rm peak} > 10^{10}M_\odot$, which is in good agreement with the number of Fornax-mass systems in the Milky Way. 

The first SMHM model (1) has a constant slope ($\alpha$) and log-scatter ($\sigma$) with halo mass:
\begin{equation}
    \mathrm{log_{10}}(M_{\rm star}) = \alpha\left(\mathrm{log}_{10}M_{\rm halo}-10\right)+6.9,  
\end{equation}
with $\sigma=\sigma_0=0.3$ dex.
The second model (2) has a halo mass-dependent slope and log-scatter \citep[see e.g.][]{munshi21}, where
\begin{equation}
\mathrm{log}(M_{\rm star}) =  \begin{cases}\alpha_1\left(\mathrm{log}_{10}M_{\rm halo}-10\right)+6.9 & \mathrm{log}_{10}M_{\rm halo} > 10  \\
                            \alpha_2\left(\mathrm{log}_{10}M_{\rm halo}-10\right)+6.9 & \mathrm{log}_{10}M_{\rm halo} \le 10, 
\end{cases}
\end{equation}

\begin{equation}
\sigma=\sigma_0+\gamma\left(\mathrm{log_{10}}M_{\rm halo}-10\right)
\end{equation}
where, $\gamma$ describes the rate at which the scatter grows. The models are chosen to roughly bracket the range of SMHM relations adopted in the literature \citep[e.g.][]{brook14, garrison-kimmel17, read17, munshi21}. However, we stress that we are not advocating a certain model, but rather exploring how different assumptions about the SMHM relation can dictate the accreted stellar mass growth of dwarf galaxies. 
Note that our adopted SMHM relations do not evolve with redshift. This assumption is motivated by the apparent lack of strong evolution found in slightly more massive systems \citep{wake11, leauthaud12,hudson15, tacchella18}, and there is little theoretical or observational evidence to suggest that we should assume otherwise for lower masses \citep[e.g.][]{fattahi20a}.

We adopt galaxy occupation models that are motivated by both theoretical and observational inferences. The parametrization of the galaxy occupation fraction follows previous work \citep{graus19,nadler20}, and is given by the following functional form
\begin{equation}
f_{\rm gal} = 0.5\left[1+\erf\left(\frac{\mathrm{log}_{10}M_{\rm halo}-\mathrm{log}_{10}M_{50}}{\sqrt{2}\sigma_{\rm gal}}\right)\right]
\label{eq:fgal}
\end{equation}
where, $M_{50}$ is the characteristic mass-scale at which 50 percent of haloes contain a galaxy, and $\sigma_{\rm gal}$ determines how quickly the galaxy occupation fraction changes as a function of halo mass. For simplicity, we adopt $\sigma_{\rm gal} =0.2$, which is admittedly chosen rather arbitrarily, but is qualitatively consistent with constraints from hydrodynamic simulations \citep{sawala16} and observations \citep{nadler20}. The (redshift-dependent) $M_{50}$ parameter is described by the following relation: $\mathrm{log_{10}}M_{50}(z) = \mathrm{log_{10}}M^{z=0}_{50}+c_1 z+c_2z^2$, where the parameters $c_1$ and $c_2$ are chosen to approximately follow the redshift dependence predicted by cosmological simulations \citep{sawala16}. 

In this work, we adopt two different galaxy occupation models (labelled ``A'' and ``B''). Model A is a fair approximation of the models given in \cite{sawala16} and \cite{benitez-llambay20}, which have a fairly high mass threshold for galaxy formation ($M^{z=0}_{50} = 10^{9.3}M_\odot$). The halo occupation fraction at $z = 0$ depends strongly on the redshift of reionization \citep[e.g.][]{bose18,benitez-llambay20, kravtsov21}. For example, the studies by \cite{sawala16} and \cite{benitez-llambay20} assume $z_{\rm re}=11.5$ \citep{planck14}.  A later epoch of reionization would result in a lower mass threshold for galaxy formation. Our second model (Model B) has a much lower threshold for galaxy formation ($M^{z=0}_{50}=10^{7.5}M_\odot$), which would perhaps be deemed unphysical given the theoretical insights described in  \cite{benitez-llambay20}. However, recent observational inferences based on the Milky Way satellite population have argued for low galaxy formation mass-thresholds \citep[e.g.][]{jethwa18, graus19, nadler20}, so it is instructive to also consider this model here. In summary, our Model A and Model B roughly span the range of galaxy occupation models, or adopted epochs of reionization (e.g. $6 \lesssim z_{\rm re} \lesssim 11$), in the literature. The parameters we use for our empirical models are given in Table \ref{tab:models}, and the forms of these galaxy models are shown in Fig. \ref{fig:gal_models}.

\subsection{Stellar mergers}
We now consider how the galaxy models described above impact the number of major and minor mergers that contribute stars to dwarf-mass haloes. First, we examine the effect of the galaxy occupation model in Fig. \ref{fig:dm_mergers_fgal}. Here, each merger event is given a probability of contributing stars to the central halo. For example, very low mass haloes will be given a low weight, as they are unlikely to host a galaxy. Note that here we are considering mergers that contribute \textit{any} stars at all - even if the stellar content is very low (i.e. 1 solar mass!). The total accreted stellar content is investigated more thoroughly in Section \ref{sec:macc}, where we find that the SMHM can have a strong influence on the total accreted stellar mass, and typical progenitor stellar masses. The black lines in Fig. \ref{fig:dm_mergers_fgal} are for $f_{\rm gal}=1$, i.e. all haloes can host galaxies, and the lines are the same as in Fig. \ref{fig:dm_mergers}. The blue lines are for Model B, which has a low mass threshold for galaxy formation ($M^{z=0}_{50} = 10^{7.5} M_\odot$). Here, the number of both major and minor mergers are reduced slightly, but the impact of these models is fairly small in \textsc{coco-cold} and negligible in \textsc{coco-warm}. On the other hand, Model A, which has a high mass threshold ($M^{z=0}_{50} = 10^{9.3} M_\odot$), significantly impacts the number of (stellar) mergers. This is particularly true for minor dark matter mergers, which only contribute $\sim 0.3-0.4$ (stellar) events per halo. The drastic reduction (by a factor of $\sim 7-16$ relative to the $f_{\rm gal}=1$ model) is because a significant number of the minor dark matter mergers in these models are ``dark'', and do not contribute any accreted stars to the central halo. Indeed, the relative importance of minor vs. major mergers for accreted stellar mass growth is in stark contrast between Models A and B: Minor mergers are dominant (by number) over major mergers when there there is a low mass threshold for galaxy formation (Model B), whereas the reverse is true when there is a high mass threshold for galaxy formation (Model A).

Note that the predicted number of stellar mergers for Model A is very similar between \textsc{coco-cold} and \textsc{coco-warm}. This is because the threshold for galaxy formation in these models roughly corresponds to the mass-scale at which \textsc{coco-cold} and \textsc{coco-warm} begin to diverge ($\sim 10^9 M_\odot$), so, in this case the haloes that typically host galaxies are actually very similar. On the other hand, there can be significant differences between \textsc{coco-cold} and \textsc{coco-warm} for Model B as here haloes can host galaxies down to lower mass scales, and this is the regime where the warm and cold dark matter subhalo properties differ.
It is also interesting to note that the relative number of (stellar) mergers predicted by different galaxy formation models are more similar in \textsc{coco-warm} relative to \textsc{coco-cold}. This is because the impact of allowing low mass haloes to form a galaxy matters less when there are very few low mass haloes around to begin with.

\subsection{Centrals vs. Satellites}
\label{sec:sats}
\begin{figure}
\centering
        \includegraphics[width=\linewidth,angle=0]{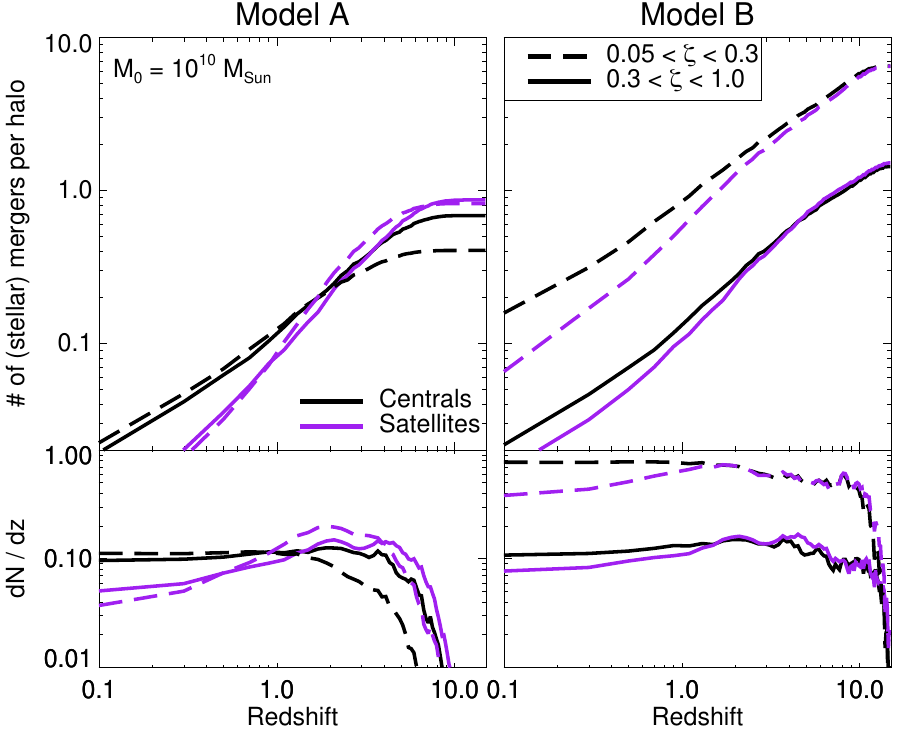}
    \caption{The number of \textit{stellar} mergers (per halo) as a function of redshift for \textsc{coco-cold}. Here, central and satellite systems at $z=0$ are shown with the black and purple lines, respectively. Each column shows a different galaxy occupation model (left-hand panel=Model A, right-hand panel=Model B).}
    \label{fig:dm_mergers_fgal_sats}
\end{figure}

Until now, we have compared different galaxy models (and dark matter models) using central haloes at $z=0$. We now consider satellite systems in the same context. In Fig. \ref{fig:dm_mergers_fgal_sats} we compare the cumulative number of stellar mergers (per halo) between central and satellite systems in \textsc{coco-cold}. The black lines are for the central haloes, which are equivalent to the lines shown in the left-hand panel of Fig. \ref{fig:dm_mergers_fgal}. The numbers of major and minor mergers for Model B (right-hand panel) are very similar between centrals and satellites. Indeed, as the halo mass threshold for this model is fairly low ($M^{z=0}_{50} =10^{7.5}M_\odot$) the merger rates are very similar to the dark matter merger rates shown in the middle panel of Fig. \ref{fig:dm_mergers}. As mentioned earlier, the satellites have lower numbers of mergers at low redshift, but slightly higher numbers at intermediate/high redshift because they reach $M_{\rm peak} \sim 10^{10}M_\odot$ at earlier times.

The difference in halo mass assembly between centrals and satellites has a much greater effect on the stellar mergers predicted by Model A (left-hand panel of Fig. \ref{fig:dm_mergers_fgal_sats}). Here, the mass threshold for galaxy formation is fairly high ($M^{z=0}_{50} =10^{9.3}M_\odot$) and has a significant influence on the number of major and minor mergers that contribute stars (see previous subsection).
At higher redshifts, the $z=0$ satellites typically have higher halo mass relative to the $z=0$ central haloes, which gives them a higher probability of accreting subhaloes that host galaxies. The larger number of mergers at higher redshift combined with the evolution of the galaxy occupation fraction, leads to a larger number of mergers that contain stars. The total number for present-day satellites is increased by a factor of 1.3(2) for major(minor) mergers relative to central haloes at $z=0$.

These findings raise an important corollary. For \textit{some} galaxy formation models (e.g. Model A), central and satellite systems with similar present-day stellar mass can have different accreted stellar mass properties (and hence stellar halos). In contrast, the accreted stellar properties will be more similar for other models (e.g. Model B). Thus, potentially, comparing the relative stellar halo properties of satellite and central systems at $z=0$ may give some insight into the underlying galaxy formation model at low mass scales. This inference is considered further in the following subsection.

\begin{figure*}
  \begin{minipage}{\linewidth}
        \centering
        \includegraphics[width=\textwidth,angle=0]{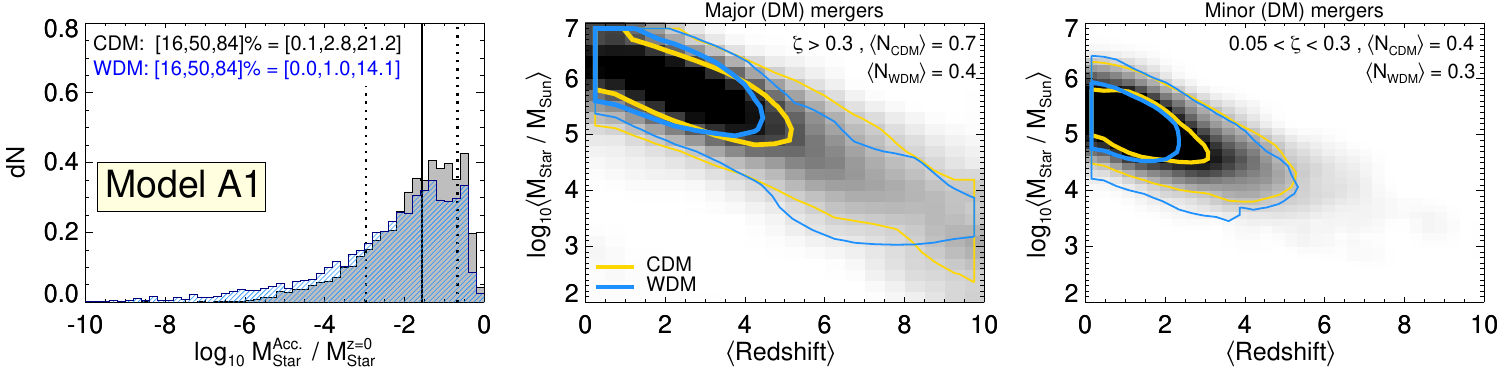}
    \end{minipage}
     \begin{minipage}{\linewidth}
       \centering
           \includegraphics[width=\textwidth,angle=0]{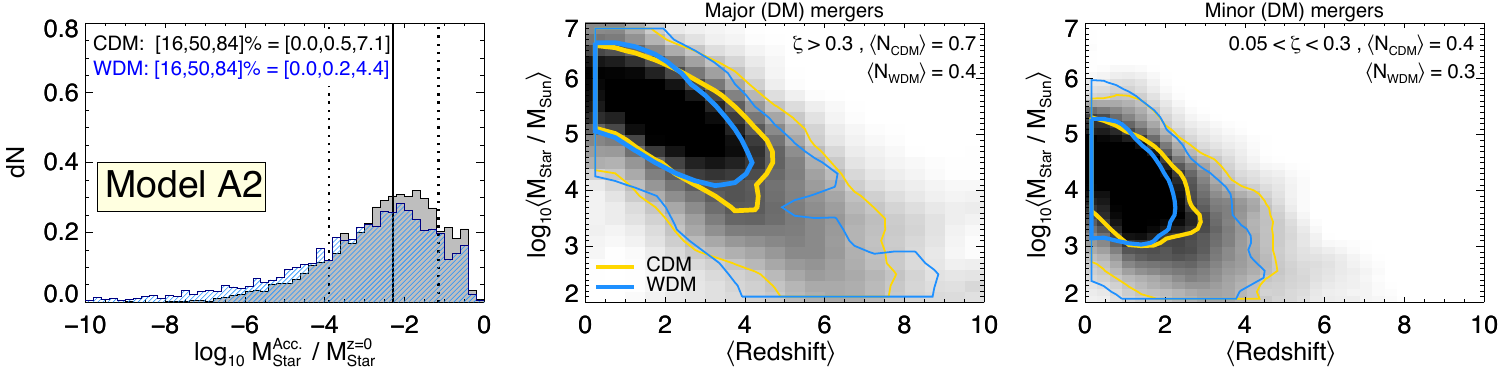}
       \end{minipage}
       \begin{minipage}{\linewidth}
        \centering
        \includegraphics[width=\textwidth,angle=0]{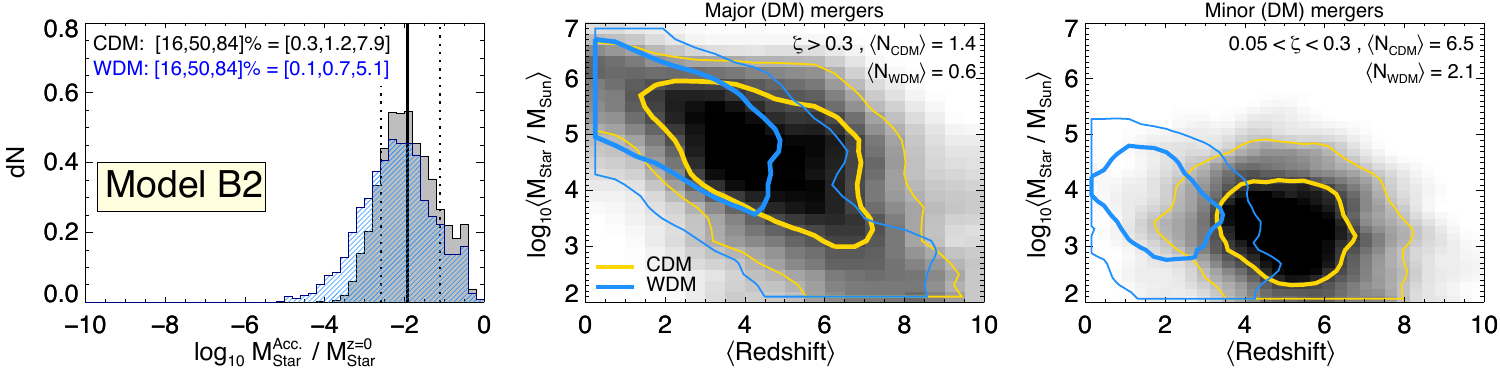}
    \end{minipage}
     \begin{minipage}{\linewidth}
       \centering
           \includegraphics[width=\textwidth,angle=0]{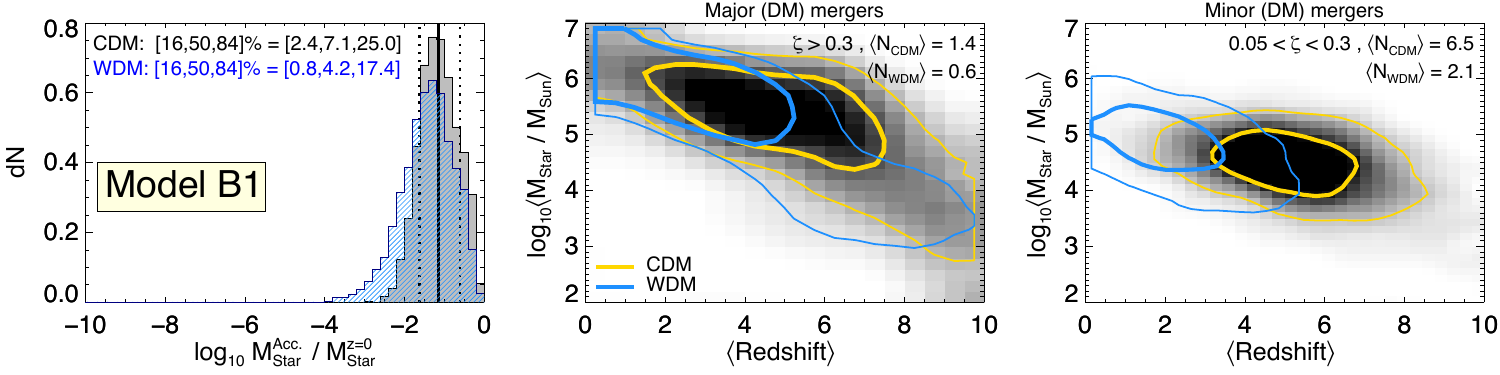}
       \end{minipage}
    \caption{The assembly of accreted stellar mass in $10^{10}M_\odot$ (z=0) haloes from \textsc{coco}. Different galaxy models are shown in each row. \textsc{coco-cold} results are shown in grayscale and gold contours, while \textsc{coco-warm} is shown in blue. \textit{Left-hand panel:} the distribution of total accreted stellar mass relative to present-day stellar mass accross all $10^{10}M_\odot$ \textsc{coco} haloes. The 16-50-84 percentiles for $100 \times M^{\rm Acc.}_{\rm Star} / M^{\rm Total}_{\rm Star}$ are given in the top left corner. \textit{Middle panel:} the (weighted) average progenitor stellar mass vs. (weighted) average redshift for major ($\zeta > 0.3$) dark matter mergers. The contours encompass 50, and 90 percent of the haloes, respectively. \textit{Right-hand panel:} the (weighted) average progenitor stellar mass vs. (weighted) average redshift for minor ($0.05 < \zeta < 0.3$) dark matter mergers. The accreted stellar mass of dwarf galaxies can vary substantially between different galaxy formation models and different types of dark matter.}
    \label{fig:macc}
\end{figure*}

\begin{figure*}
  \begin{minipage}{\linewidth}
        \centering
        \includegraphics[width=\textwidth,angle=0]{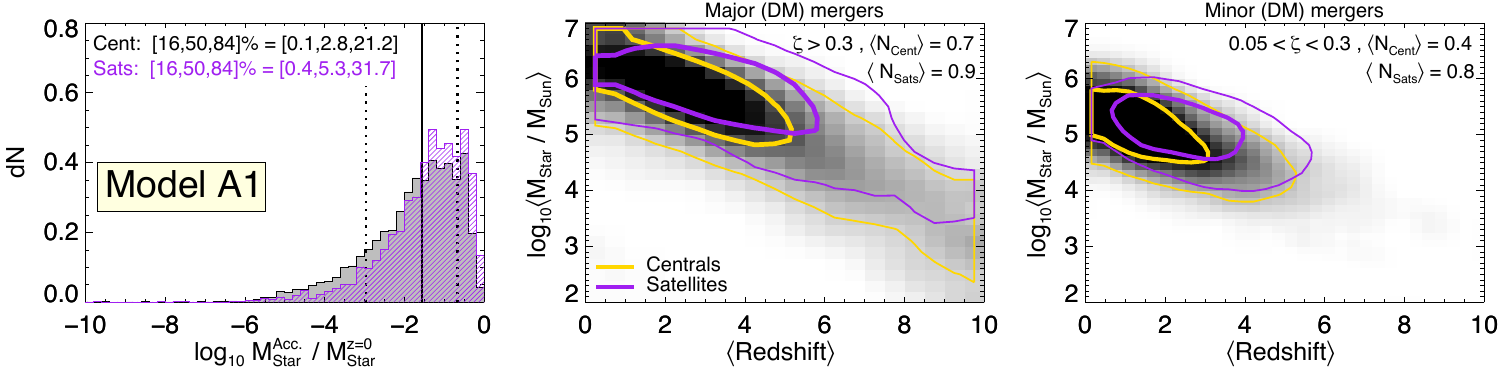}
    \end{minipage}
     \begin{minipage}{\linewidth}
       \centering
           \includegraphics[width=\textwidth,angle=0]{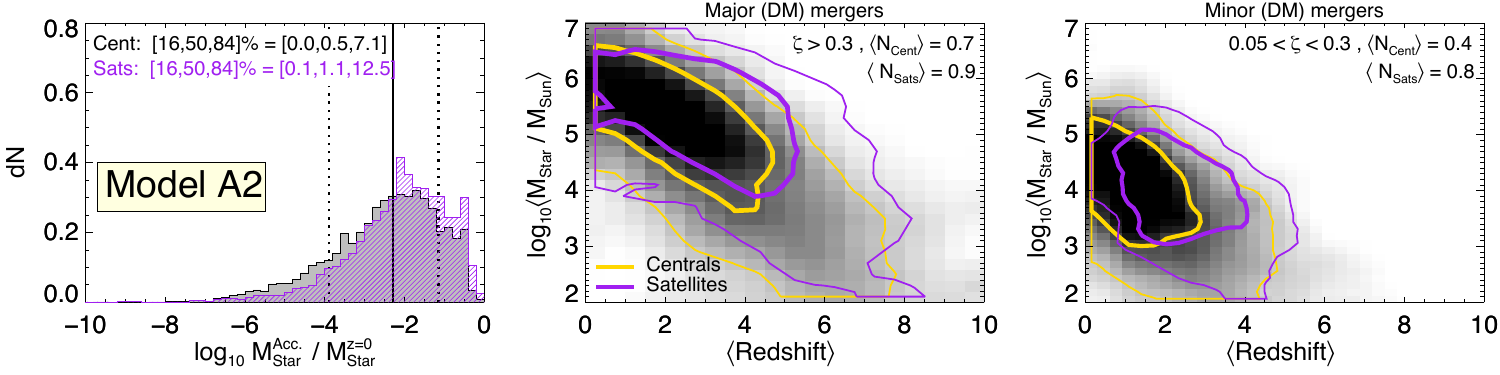}
       \end{minipage}
       \begin{minipage}{\linewidth}
        \centering
        \includegraphics[width=\textwidth,angle=0]{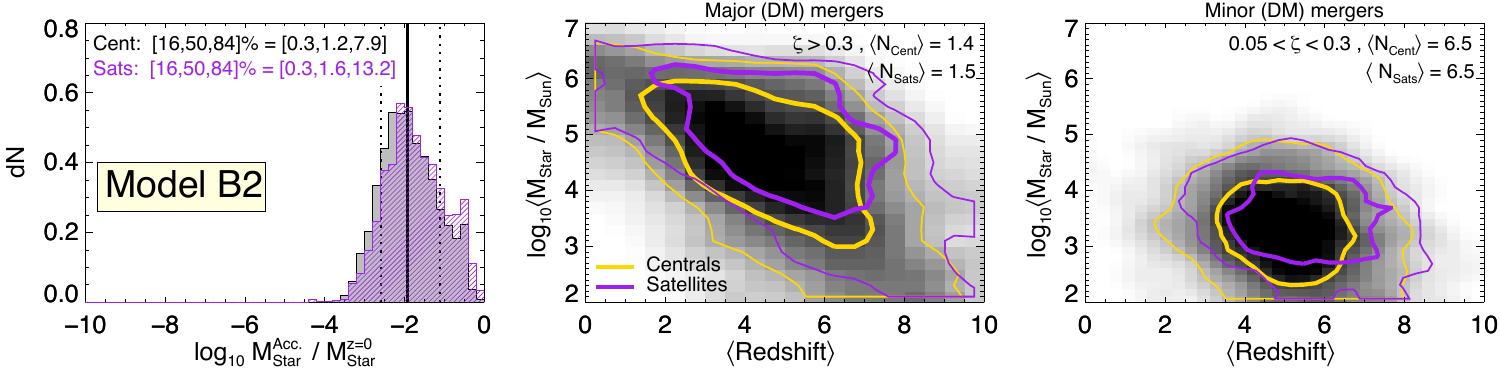}
    \end{minipage}
     \begin{minipage}{\linewidth}
       \centering
           \includegraphics[width=\textwidth,angle=0]{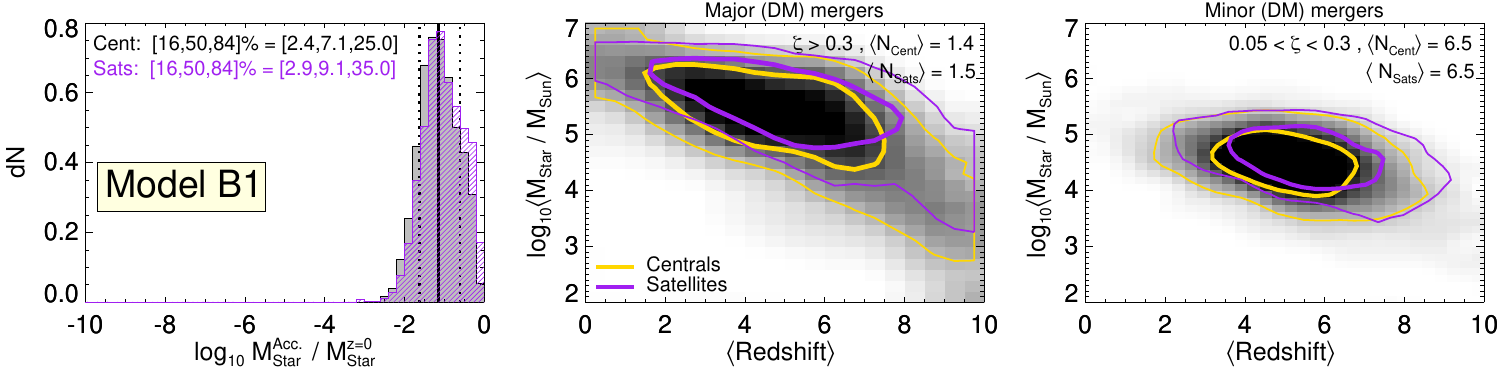}
       \end{minipage}
    \caption{Same as Fig. \ref{fig:macc}, but comparing central (greyscale, gold contours) and satellite (purple) haloes at $z=0$ in \textsc{coco-cold}. }
    \label{fig:macc_sats}
\end{figure*}

\subsection{Accreted stellar mass}
\label{sec:macc}
We now incorporate the full galaxy models and apply both galaxy occupation probabilities and a SMHM relation. Here, as well as having a galaxy occupation probability, the dark matter haloes are assigned a stellar mass according to the SMHM relation. Note that as we include scatter in the SMHM relations we use 1000 Monte Carlo trials to compute the stellar mass associated with dark matter mergers (assuming the scatter in stellar mass at fixed halo mass is normally distributed in log-space). In Fig. \ref{fig:macc} we consider the assembly of accreted stellar mass in these (central) dwarf-mass ($10^{10}M_\odot$) haloes, where different galaxy models are shown in each row. Results for both \textsc{coco-cold} (greyscale, gold contours) and \textsc{coco-warm} (blue) are shown.

The left-hand panels of Fig. \ref{fig:macc} show the total accreted stellar mass relative to the stellar mass of the central galaxy (at $z=0$). This is computed using the SMHM relations (Models 1 and 2) to calculate stellar masses for each halo, and $f_{\rm gal}$ (Models A and B) gives the probability of a subhalo hosting a galaxy: $M^{\rm Acc.}_{\rm Star} = \sum_i f_{\rm gal,i} M_{\rm Star, i}$, where $f_{\rm gal,i}$ and $M_{\rm Star, i}$ are the occupation probabilities and stellar masses for each destroyed subhalo. The total accreted stellar mass can vary significantly, owing both to the galaxy formation models and the intrinsic halo-to-halo scatter. However, it is worth noting that the models with high mass threshold for galaxy formation are less likely to have a discernible stellar halo: for example, in \textsc{coco-cold} $\approx$ 15-30\% of the haloes have accreted stellar mass $M^{\rm Acc.}_{\rm Star} / M^{\rm Total}_{\rm Star} \lesssim 0.1\%$ in the high mass threshold models. In comparison, only $\lesssim$ 5\% of the haloes have such small accreted stellar masses in the low mass threshold models. In addition, models with a shallower SMHM slope tend to have larger accreted stellar masses (by a factor of $\sim 6$). Thus, these simple empirical models predict that the contribution of accreted stars to dwarf galaxy haloes (i.e. their stellar haloes) is strongly dependent on the galaxy formation model. Indeed, even the mere \textit{detection} of a stellar halo surrounding a dwarf galaxy can likely tell us something about galaxy formation at low mass scales. The total amount of accreted stellar mass is also dependent on the dark matter model; there is typically less accreted stellar material (by a factor of $\sim 2$) in \textsc{coco-warm} relative to \textsc{coco-cold}.

The middle and right-hand panels of Fig. \ref{fig:macc} show the typical progenitor stellar mass vs. average redshift for major (middle), and minor (right) dark matter mergers.
 Here, we show the weighted average values for each halo, where the ``weight'' is given by $f_{\rm gal}$, e.g. $\langle x \rangle = \sum_i x_i f_{\rm gal, i} / \sum f_{\rm gal, i}$.  The models with high halo mass threshold for galaxy formation (Models A1 and A2), typically have a low number ($\langle N \rangle \leq 1$) of relatively massive accretion events (log$_{10}\langle M_{\rm star}/M_\odot \rangle \sim 5-6$) at low/intermediate redshift ($z \sim 0-4$). The models with low mass threshold tend to have more ($\langle N \rangle \sim 6$) low mass mergers (log$_{10}\langle M_{\rm star}/M_\odot \rangle \sim 3-5$) at higher redshifts ($z \sim 3-6$). In addition, the typical progenitor stellar masses, and hence the overall accreted stellar mass, are higher for models with shallower SMHM slopes (Models A1, B1). This exercise shows that the typical contributors to the accreted components of dwarf galaxies can vary considerably between different galaxy models. In particular, the galaxy occupation model has a big influence, as this essentially dictates whether or not minor (stellar) mergers can occur at the dwarf mass scales. This is an important point, as the phase-space distribution of accreted stars can vary substantially between major and minor mergers. We explore this more thoroughly in the following section.

As mentioned earlier, \textsc{coco-cold} and \textsc{coco-warm} give very similar predictions for galaxy models with a high threshold for galaxy formation (Models A1 and A2). Here, the haloes that host galaxies are typically above the mass-scale at which we see differences between the two dark matter models. However, for galaxy models with a low threshold for galaxy formation (Models B1 and B2) we see important differences between \textsc{coco-cold} and \textsc{coco-warm}. The number of major and minor mergers are reduced by factors of 2 and 3 in \textsc{coco-warm}, respectively. In addition, the mergers in \textsc{coco-warm} typically occur at lower redshift than \textsc{coco-cold}. This is likely because the formation times of low mass haloes in \textsc{coco-warm} are delayed relative to \textsc{coco-cold} \citep[see][Figure 4]{bose16}.

In Fig. \ref{fig:macc}, we compared the accreted stellar properties for different galaxy formation and dark matter models for dwarf galaxies at $z=0$ that are centrals. We now consider how satellite systems at $z=0$, which have similar $M_{\rm peak}$ values, compare. Here, we only show \textsc{coco-cold} but the relative differences we find can also be seen in the \textsc{coco-warm} model. The greyscale and gold contours in Fig. \ref{fig:macc_sats} are for the central haloes (the same as in Fig. \ref{fig:macc}), and the purple shades and contours are for the satellites. In general, for each galaxy formation model, the total accreted stellar mass is higher for the satellite systems.  However, this difference is especially prominent for Models A1 and A2. Here, the fact that the satellites have higher halo mass at higher redshift enables more mergers to occur that include stars. The middle and right-hand panels show that both minor and major mergers tend to happen at higher redshift for the $z=0$ satellites. Again, the differences are more subtle for Models B1 and B2, because here most merger events are able to contribute accreted stars. However, this bias in redshift enables more major and minor mergers that contribute stars to occur in Models A1 and A2. The difference in minor mergers may be particularly important. For central haloes at $z=0$ the contribution of minor mergers to the accreted stellar mass is fairly insignificant, and the chance of finding cold features in the stellar halo of these dwarfs seems unlikely. However, the satellite systems are twice as likely to experience minor mergers that create stellar streams. Importantly, the relative difference between the stellar haloes of centrals and satellites may be an important discriminator between different galaxy formation models.

\section{Dwarf-dwarf mergers}
\label{sec:gadget}
In this section, we explore how merger events influence the build-up of dwarf galaxy stellar haloes. To this end, we use a suite of isolated models of merger events between dwarf haloes. Here, we only consider dry mergers, and employ a dark matter ``tagging'' technique to model the stellar components embedded within the dark matter haloes (cf. \citealt{cooper10}).

\subsection{Gadget simulations}
We use the publicly available \textsc{Gadget-2} code to simulate collisionless mergers between satellites and centrals. Here, we assume that the dark matter haloes have spherically symmetric Navarro-Frenk-White (NFW, \citealt{nfw}) density distributions.  Throughout, we adopt a halo concentration $c_{200}=r_{200}/r_s=10$, which is appropriate for (CDM) dwarf galaxies at low/intermediate redshift. We check that our results are not significantly changed if we adopt different halo concentrations (i.e. similar qualitative results are seen for haloes with $c_{200}=5$ and $20$). This is important because the typical halo concentrations can change with redshift and varies between WDM and CDM \citep[e.g.][]{ludlow16}. Thus, to first order, our illustrative results are relevant for both the CDM and WDM cases we consider in this work, but a more detailed analysis would require more realistic halo profiles. All haloes (satellites and centrals) are generated with $N=10^6$  particles, and the initial conditions are generated\footnote{Initial conditions are generated using the ICICLE python package \citep{drakos17}.} using an exponentially truncated NFW profile \citep{springel99}, with decay parameter $d=10r_s$ \citep[see e.g.][]{penarrubia10}. After initializing the particles in phase space, each halo is run in isolation for $T=2$ Gyr to enable it to dynamically relax before any merger events occur.

We consider three different central halo masses\footnote{Here, $M_{200}$ is the mass contained within the radius, $r_{200}$, which encloses a mean density equal to 200 times the critical density of the universe.}, $M_{200}=10^{10},10^{9.5}, 10^{9} M_\odot$, which roughly represent a present-day $10^{10}M_\odot$ mass halo at $z=0$, $2,$ and $4$, respectively. We adopt force softenings of $\epsilon=50, 35, 25$ pc for each central halo mass. Using the relation derived in \cite{ludlow19}, the convergence radius is given by: $r_{\rm conv} = 1.1 r_{200}/N_{200}^{0.4}$, where $N_{200}$ gives the number of particles within the virial radius, $r_{200}$. Thus, in our simulation runs, $r_{\rm conv} \sim 4 \times \epsilon$. Given that we are mainly interested in the outer regions of the dwarf galaxies, we find that our results are not significantly affected by changes in the particle number and/or force softening.

For each central, we simulate satellite mergers with 14 equally spaced log-mass ratios in the range, $-1.3 < \mathrm{log}_{10}\zeta < 0$. Note that this mass-ratio is the key dynamical quantity, which will shape where the accreted material will end up (see below). The initial orbits of the satellites are described by their energy and angular momentum. We assume that the orbital energy of the satellites at infall is equivalent to the energy of a circular orbit at the virial radius of the host halo, i.e. $E_{\rm inf} = E_{\rm circ}(r_{\rm circ})$, where $r_{\rm circ} \approx c_{200} r_s = r_{200}$. This assumption is motivated by the work by \cite{jiang15} (also see \citealt{amorisco17}), who show that this relation is approximately true for satellite orbits in cosmological simulations. Finally, for the initial angular momentum, we adopt a circularity, $j=J/J_{\rm circ}(E) =0.5$, which is approximately the median value found in cosmological simulations \citep[e.g][]{benson05, jiang15}. Note that we also consider other values of circularity ($j=0.2$ and $0.8$) for comparison, and find that our results are not significantly changed.

All simulations are run for $T=15$ Gyr to ensure the mergers are completed, and we analyse the outputs at the end of the simulation. In the following subsection, we describe how we assign stars to these dark matter mergers.

\begin{figure}
  \begin{minipage}{0.5\linewidth}
        \centering
        \includegraphics[width=\textwidth,angle=0]{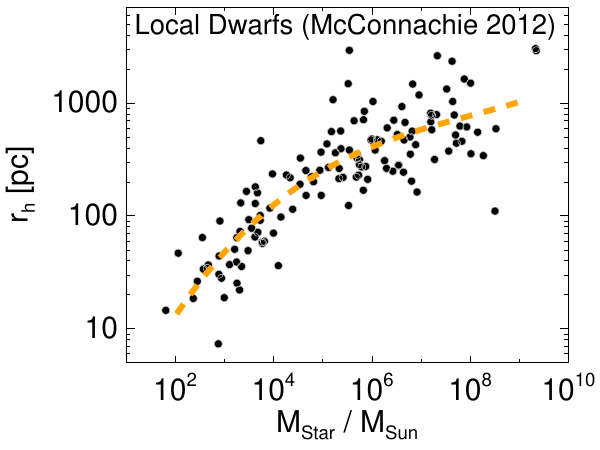}
    \end{minipage}
     \begin{minipage}{0.5\linewidth}
       \centering
           \includegraphics[width=\textwidth,angle=0]{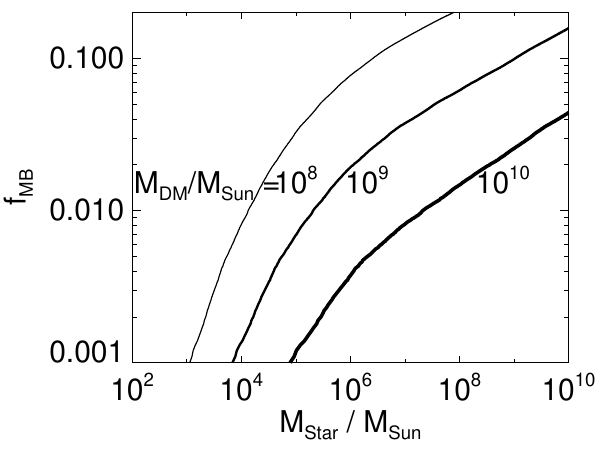}
       \end{minipage}
    \caption[]{\textit{Left-hand panel:} the half mass radius vs. stellar mass for Local Group dwarfs (data from \cite{mcconnachie12}. The orange dashed line is a polynomial fit to the relation, which we use to tag dark matter particles for any given stellar mass. \textit{Right-hand panel:} the fraction of most bound dark matter particles that are tagged as ``stars'' as a function of stellar mass. Each line shows the relation for different mass haloes. This procedure ensures that the tagged stellar component approximately has the correct spatial distribution. }
    \label{fig:tag}
\end{figure}
    
\subsection{Dark matter tagging}
The setup we have described above follows an $N-$body collision between two dark matter haloes. We now discuss how we assign stars to these dark matter-only simulations. A common technique that has been used in the literature is ``dark matter tagging'', whereby a certain fraction of the most bound dark matter particles are tagged as stars \citep[e.g.][]{delucia08, cooper10}. It has been shown that such dark matter tagging schemes result in roughly exponential profiles, and give a reasonable approximation to the true stellar density profiles of dwarf galaxies \citep[e.g.][]{cooper17, lebret17}. The advantage of this method is that the tagging is performed after the simulations have run, so, in principle, the stellar component assigned to each halo can be changed \textit{ad hoc}. This is particularly useful when considering SMHM relations with significant scatter.

Our approach is to consider a particle tagging method that approximately represents the \textit{true} sizes of known dwarf galaxies. In the left-hand panel of Fig. \ref{fig:tag} we show the relation between half-light radius and stellar mass for a compilation of local dwarf galaxies \citep{mcconnachie12}. The dashed orange line shows a polynomial fit to the relation, and we use this to inform our particle tagging method. Essentially, for a given stellar mass, we tag the fraction of most bound dark matter particles, $f_{\rm MB}$, that is able to reproduce this size-mass relation. The right-hand panel of Fig. \ref{fig:tag} shows the resulting relation between $f_{\rm MB}$ and stellar mass for a range of halo masses. Note that we ensure that each stellar component is represented with at least 100 particles, which sets a lower limit of $f_{\rm MB} = 10^{-4}$ (as our simulations are run with $N=10^6$ particles). Thus, the spatial distributions of very low stellar mass dwarfs ($M_{\rm star} \lesssim 10^3M_\odot$) embedded in massive haloes are only very approximate, as we cannot properly resolve these systems. However, we find that, while the inner spatial distributions of such cases are poorly resolved, this limitation does not significantly affect our analysis of the outer halo properties.

This tagging method has significant limitations, and the evolution of half-mass radius during/after merger events can be non-trivial. This is especially true as we do not take into account gas in our simulations, which can have an important effect on the sizes of dwarf galaxies, particularly for major mergers at early times \citep[e.g.][]{benitez-llambay16, tarumi21}. However, for our purposes, we are mainly interested in the stellar distribution in the very outskirts of dwarf haloes, and we only use the half-light (or half-mass) radius in our tagging approach. We consider the ``stellar halo'' component as material beyond $\approx 0.1r_{200}$, which is typically $5-20 \times$ further out than the half-light radius.

\subsection{Dwarf stellar haloes}
\begin{figure}
  \begin{minipage}{\linewidth}
        \centering
        \includegraphics[width=\textwidth,angle=0]{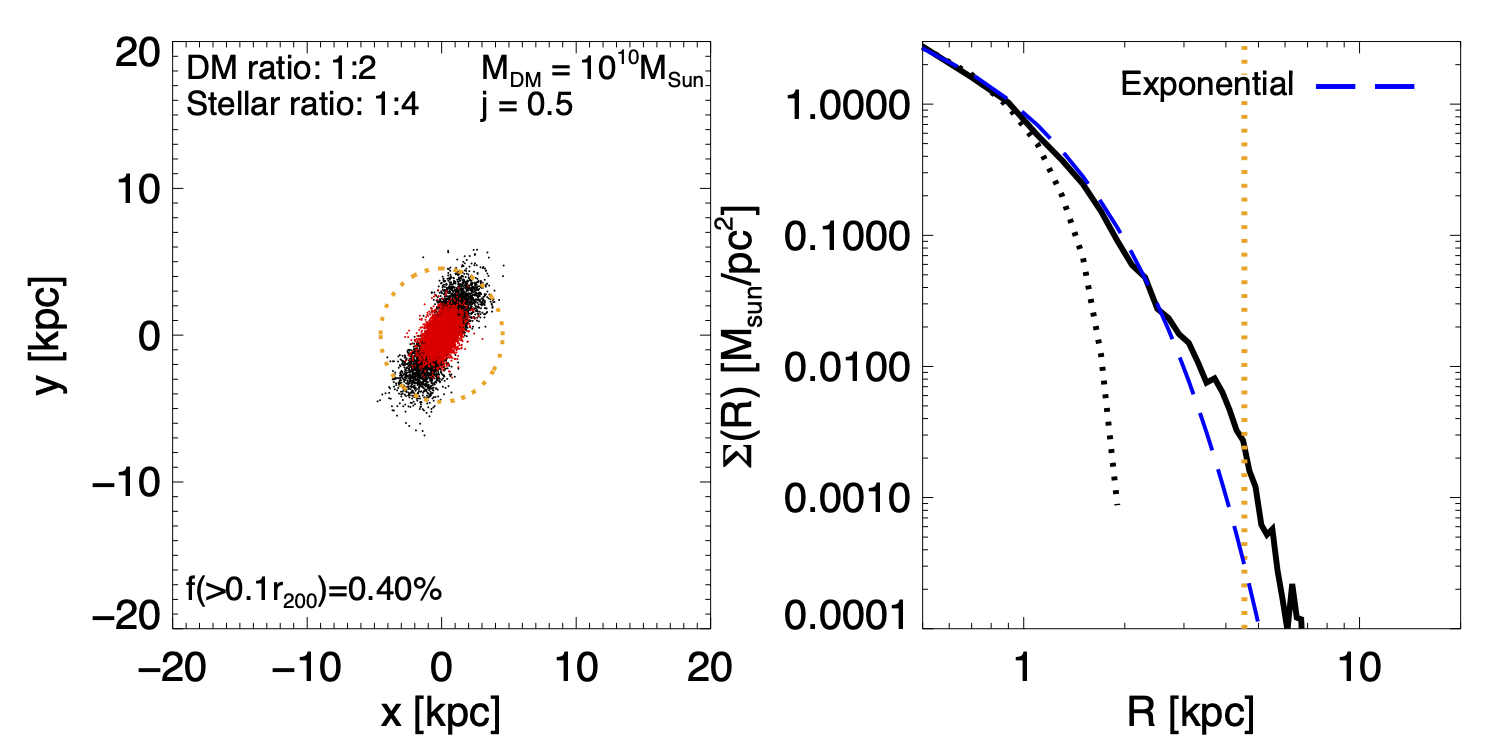}
    \end{minipage}
     \begin{minipage}{\linewidth}
       \centering
           \includegraphics[width=\textwidth,angle=0]{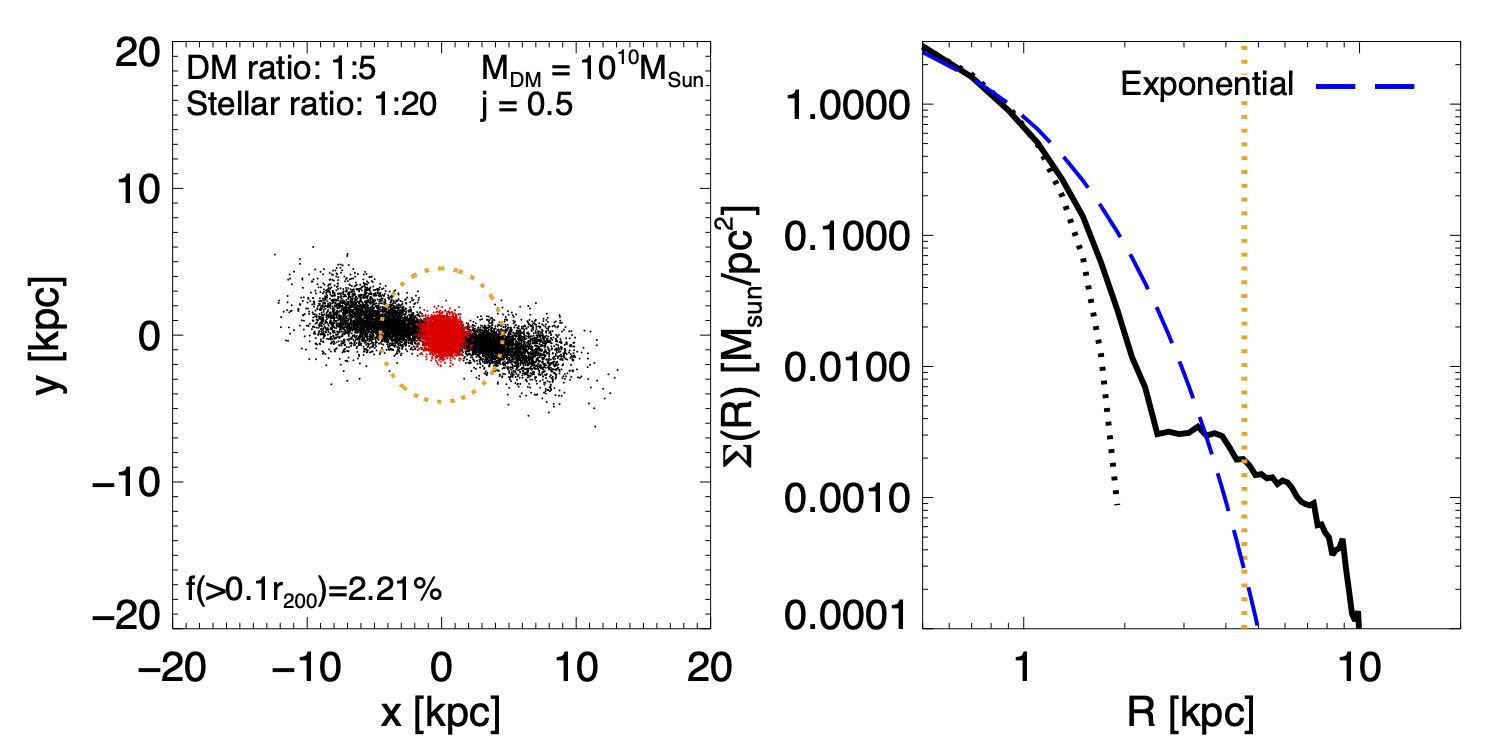}
       \end{minipage}
       \begin{minipage}{\linewidth}
       \centering
           \includegraphics[width=\textwidth,angle=0]{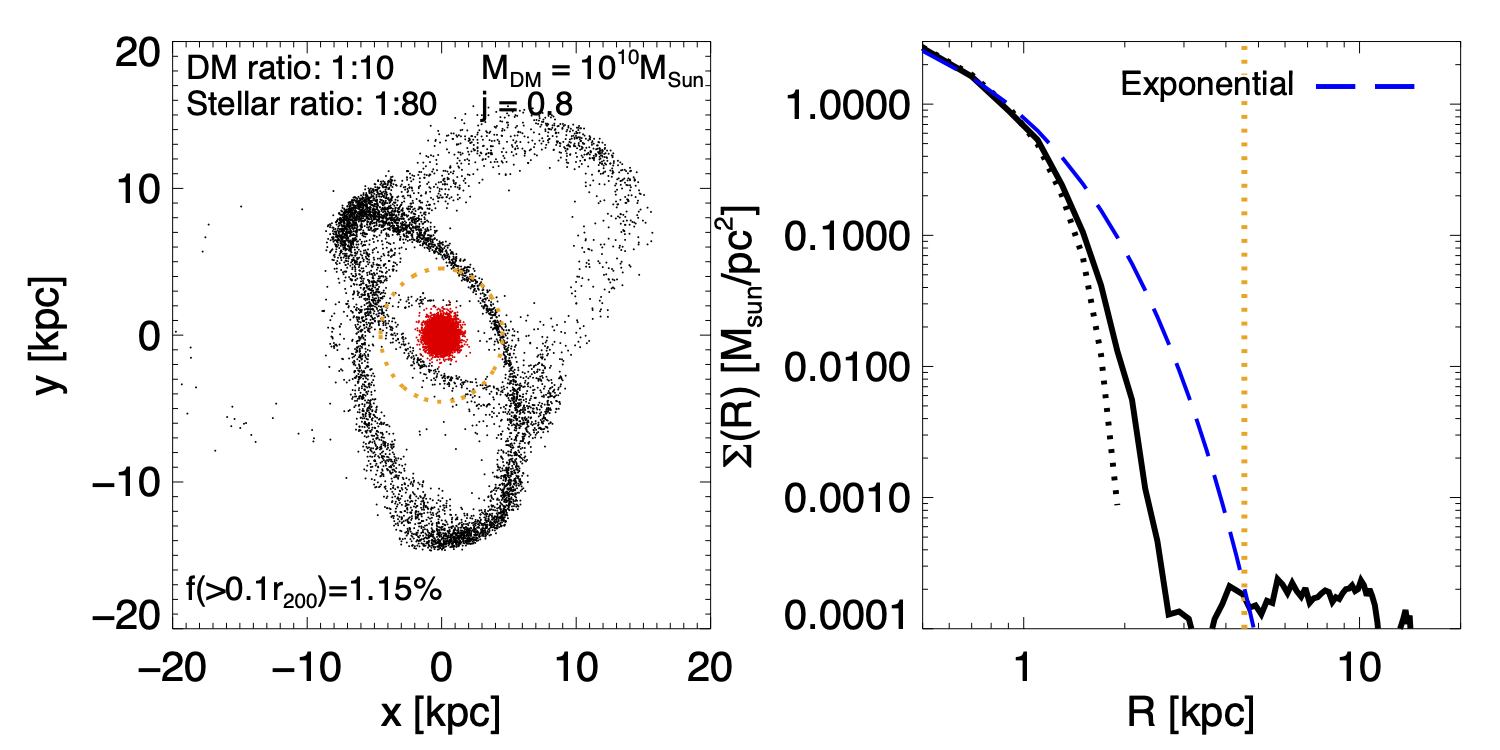}
       \end{minipage}
    \caption{Examples of dwarf-dwarf mergers. The left-hand panels show the distribution of stars in the $x-y$ plane. The red points are the central, and the black points the destroyed satellite. The right-hand panel shows the projected radial distribution of the stars. The dotted black line shows the initial profile of the central galaxy, and the blue dashed line shows a fit to an exponential profile. Major mergers can cause the stellar distribution to extend, but owing to the large binding energy have little material at very large distances in the halo.}
    \label{fig:examples}
\end{figure}
In Fig. \ref{fig:examples}, we show some examples of the resulting stellar distribution from different dark matter mass merger ratios $T=15$ Gyr after infall for $10^{10}M_\odot$ centrals. The left-hand panels show the projected distribution in the $x,y$ plane (red = central, black = satellite), and the right-hand panel shows the surface density profile. The top panel shows a major merger (1:2 in dark matter), the middle panel shows an ``intermediate'' merger (1:5 in dark matter), and the bottom panel shows a minor merger (1:10 in dark matter). This figure highlights an important difference between major and minor mergers. For major mergers, the central galaxy can be ``heated'' and the combination of the central and satellite stellar material forms an extended distribution. However, the fraction of stellar material beyond $\sim 0.1 r_{200}$ (relative to the total stellar mass) is fairly small (0.4 \% in the example given in the top panel). This is because the binding energy of the satellite merger is relatively high. On the other hand, for minor mergers the stellar component of the central is hardly affected, but, as the satellite has a lower binding energy, stripped material from the satellite can be splayed out to larger distances from the central galaxy. Thus, the amount of stellar material at large distances in the halo is largely determined by two competing effects: (1) the amount of stars brought in by the satellite galaxy, and (2) the binding energy of the infalling satellite.  In addition, dynamical friction plays a key role, as typically major mergers decay and \textit{then} get stripped, whereas minor mergers get stripped before they can decay, and are thus more populous in the halo outskirts. Note that these findings are not restricted to dwarf mass-scales. In particular, the dynamical influence of major vs. minor mergers were discussed in detail in \cite{amorisco17}. Indeed, it is the mass-ratio rather than the mass scale that drives the relative differences between the two merger processes.

\begin{figure}
  \begin{minipage}{\linewidth}
        \centering
        \includegraphics[width=\textwidth,angle=0]{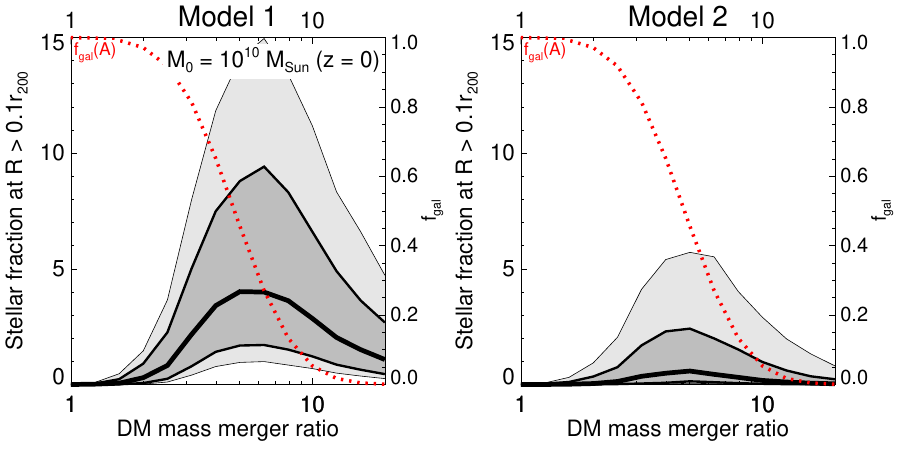}
    \end{minipage}
     \begin{minipage}{\linewidth}
       \centering
           \includegraphics[width=\textwidth,angle=0]{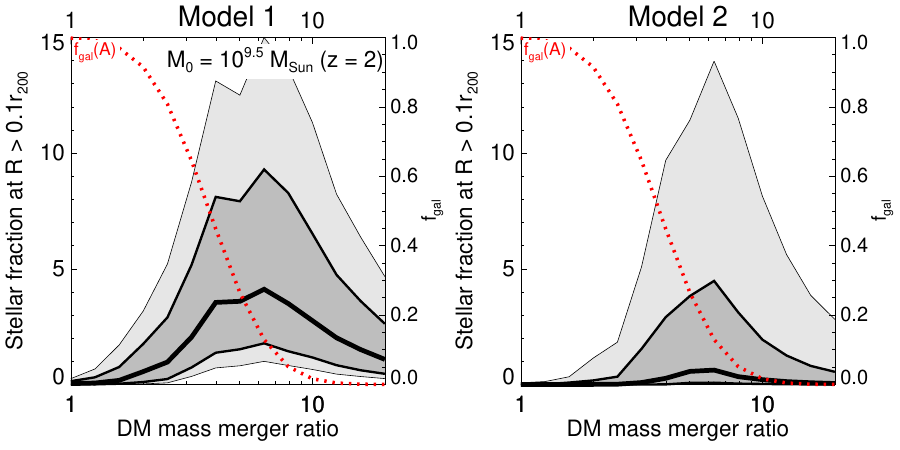}
       \end{minipage}
       \begin{minipage}{\linewidth}
       \centering
           \includegraphics[width=\textwidth,angle=0]{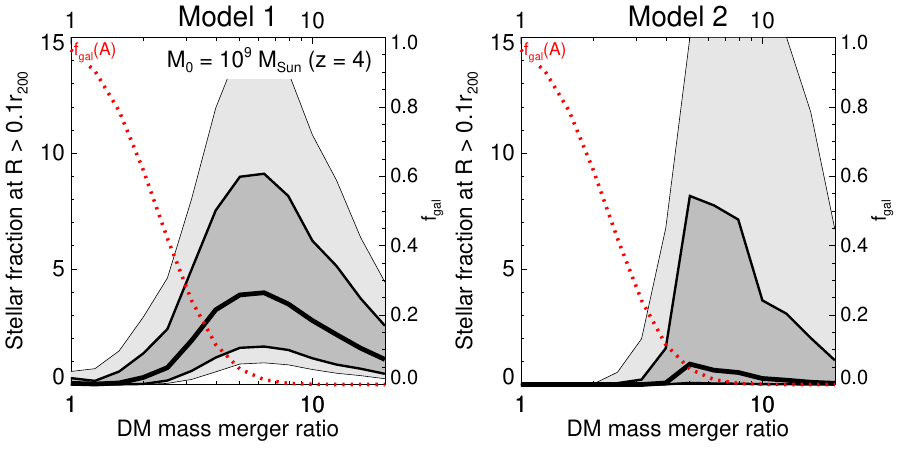}
       \end{minipage}
       \caption{The (percentage) fraction of stellar material at large distances in the halo ($> 0.1r_{200}$) as a function of dark matter merger ratio ($1/\zeta$). The thick black line indicates the median, and the dark and light grey shaded regions show the 68 and 95 percentiles, respectively. Each panel shows a different mass central halo at different redshift (roughly chosen to mimic the mass evolution of a present-day $10^{10}M_\odot$ halo). The columns indicate two different stellar mass-halo mass relations. Models with a shallower stellar mass-halo mass relation tend to have larger fractions of stars at large distances in the halo. Interestingly, the stellar material at large distances is maximised for dark matter merger ratios of $\sim$ 1:5. Here, the binding energy is not too high (cf. major mergers), but there is still significant stellar material brought in (cf. minor mergers). The red dashed line shows the galaxy occupation fraction for a high mass threshold model (Model A - note Model B has approximately $f_{\rm gal}=1$ in all these figures). Here, one can see that, regardless of the SMHM relation, minor mergers are largely ineffective at bringing in stars, which limits the amount of stellar mass that can be found at large distances in the halo.}
    \label{fig:frac_star}
\end{figure}
The fraction of stellar material at large distances (relative to total stellar mass) is explored more thoroughly in Fig. \ref{fig:frac_star}. Here, we show the stellar fraction at (projected) distances greater than $0.1r_{200}$ as a function of dark matter mass ratio. Each column shows different SMHM models, and each row has a different mass central halo. The scatter in the SMHM relation is included by running $N=5000$ Monte Carlo trials. Models with a relatively shallow SMHM slope at low mass scales have larger stellar halo fractions, this is because they typically assign larger stellar masses for a given halo mass. However, when there is significant scatter in the SMHM relation, even models with a fairly steep SMHM slope can sometimes lead to cases with very high stellar fractions. This can be seen in the right-hand panels of Fig. \ref{fig:frac_star}; here, the median values are fairly low, but the large scatter in the model allows for very high fractions within the 95 percent confidence regions. Moreover, the scatter increases for lower halo masses because the SMHM relation for this model (Model 2) has a growing scatter with decreasing halo mass (see Fig. \ref{fig:gal_models}).

\begin{figure*}
    \begin{minipage}{0.48\linewidth}
	\includegraphics[width=\textwidth]{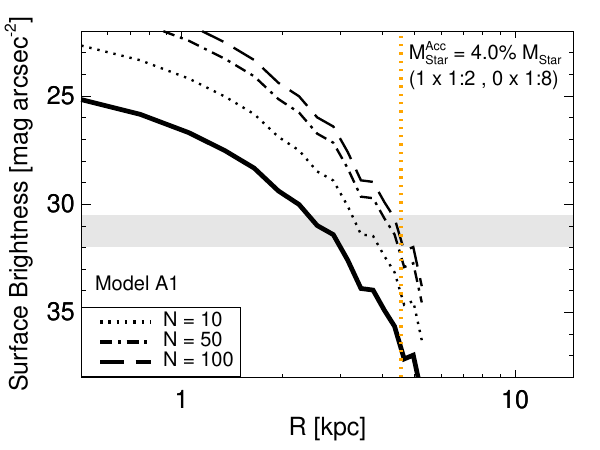}
	\end{minipage}
	\begin{minipage}{0.48\linewidth}
	\includegraphics[width=\textwidth]{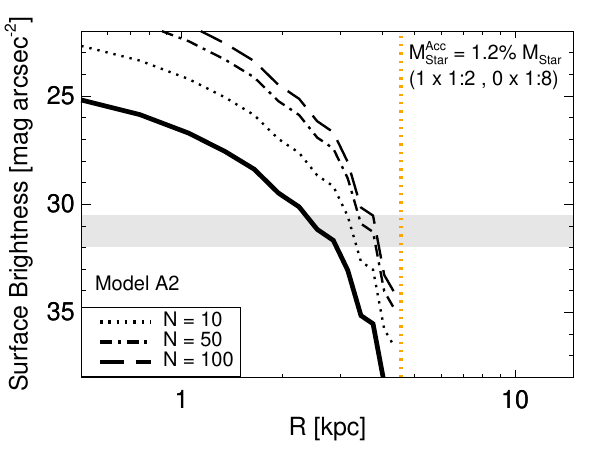}
	\end{minipage}
	\begin{minipage}{0.48\linewidth}
	\includegraphics[width=\textwidth]{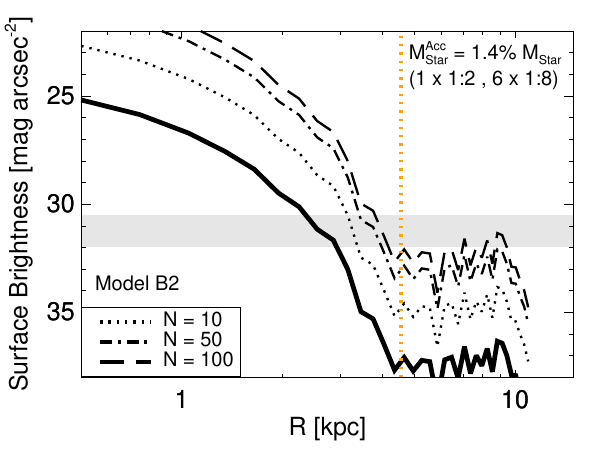}
	\end{minipage}
	\begin{minipage}{0.48\linewidth}
	\includegraphics[width=\textwidth]{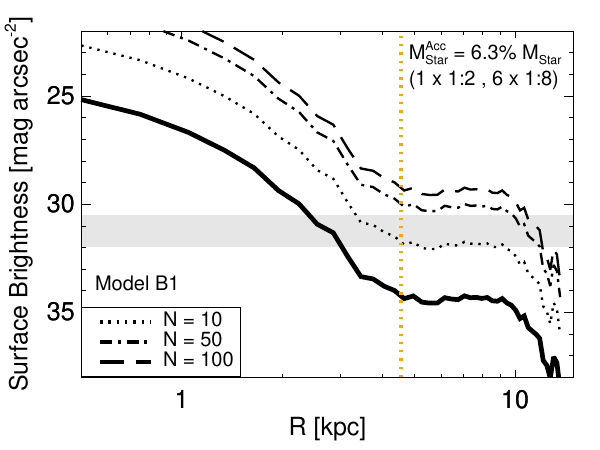}
	\end{minipage}
    \caption{Example surface brightness profiles of $10^{10}M_\odot$ mass dwarfs. Each panel shows an illustrative example for a given galaxy model. To construct these, we adopt the typical dark matter mass ratio, accreted stellar mass and number of mergers for each empirical galaxy model. The thick solid line is the average profile for an individual dwarf, and the different line styles indicate the profile if different numbers of dwarfs are stacked together. The dotted orange line indicates 0.1$r_{200}$ of the central. The shaded gray band shows the approximate surface brightness limit of the Vera Rubin Observatory LSST \citep{brough20}. In order to detect these features several dwarf galaxies would need to be stacked, and only Model B1 would likely produce a detectable stellar halo at these mass-scales.}
    \label{fig:surf_brightness}
\end{figure*}

For both SMHM models the stellar halo fraction is \textit{maximised} when the dark matter mass ratio is $\approx$ 1:4 to 1:8. This is because these ``intermediate'' mass-ratio cases do not have such high binding energy that material is confined to the inner regions (cf. major mergers), but they can still have have enough stellar mass to make a significant contribution (cf. minor mergers).  We also show in Fig. \ref{fig:frac_star} the galaxy occupation fraction with the red dotted line. Note this is not shown for Models B1 and B2 because here $f_{\rm gal} \approx 1$ in all cases. The inclusion of the galaxy occupation models highlights another important point: for models with a high halo mass threshold for galaxy formation, it is typically only \textit{major} mergers that can contribute accreted stars.

 We have found that, owing to the high binding energy, major mergers do not contribute many stars to to the outer halo ($\gtrsim 0.1 r_{200}$). However, we are neglecting an important effect --- gas rich mergers. Previous work has shown that mergers involving gas-rich dwarfs can have a considerable affect on the star formation and stellar distribution of the merger remnant. For example, \cite{benitez-llambay16} (see also \citealt{genina19}) show how major (wet) dwarf mergers can form the ``two-halo'' populations seen in classical dwarf galaxies: i.e. a centrally concentrated young component and a more extended older component. This extended old component is akin to the radially extended populations we have seen from the dry mergers in this work. In addition, \cite{tarumi21} showed that the amount of stellar material dispersed to large radii after a major merger can be enhanced when the merger is wet. These findings suggest that we are likely underestimating the impact of major mergers on the (distant) stellar halo. However, it is worth emphasizing that all of the galaxy models we are considering predict some major mergers, albeit there is a difference in number by a factor of 2-3 between different models. On the other hand, the difference in the number of minor mergers is much more impactful: here, the numbers can differ by a factor of 20. Thus, searching for signatures of minor mergers (e.g. tidal features, cold stellar streams, very low metallicity stars) at dwarf mass scales will likely provide the biggest discriminator between different galaxy formation models.
 
\section{Observing dwarf stellar haloes}
\label{sec:obs}
 We now discuss the observability of the predicted dwarf galaxy stellar haloes based on our $N$ body simulations and empirical galaxy models. First, we consider the surface brightness levels of the dwarf stellar haloes. To do this, we adopt typical merger event characteristics from \textsc{coco-cold} for each empirical galaxy model (see Section \ref{sec:gals}), and apply this to our isolated \textsc{Gadget} simulations. For Models A1 and A2, a $10^{10}M_\odot$ field dwarf today has typically experienced $N \leq 1$ major mergers, and no, or very few, minor mergers. For these cases, we assume 1 major merger with mass-ratio 1:2, and adopt the progenitor masses of $M_{\rm star}=10^{5.5}M_\odot$ and $M_{\rm star}=10^{5.0}M_\odot$, respectively (see Fig. \ref{fig:macc}). In Models B1 and B2, there is typically $N \sim 1$ major merger, and $N \sim 6$ minor mergers. For these, we also assume 1 major merger with dark matter mass-ratio 1:2, and adopt the progenitor masses of $M_{\rm star}=10^{5.0}M_\odot$ and $M_{\rm star}=10^{5.5}M_\odot$, respectively. However, we also include the 6 minor mergers; assuming 1:8 dark matter ratios and progenitor masses of $M_{\rm star}=10^{3.4}M_\odot$ and $M_{\rm star}=10^{4.5}M_\odot$, respectively. 

The resulting surface brightness profiles are shown in Fig. \ref{fig:surf_brightness}. Here, we have assumed a mass-to-light ratio of 1 in order to convert surface mass density to surface brightness. As well as a typical individual halo profile, we have also indicated stacked profiles with the different line styles. It is clear that for any of the models the stellar halo of an \textit{individual} dwarf galaxy will be extremely difficult, if not impossible, to detect. However, there is some hope if several dwarf galaxies can be stacked together. Importantly, the only model that would likely produce stellar haloes detectable by the Vera Rubin Observatory Legacy Survey of Space and Time \citep[LSST][]{brough20} is Model B1. This model has a fairly shallow faint end slope to the SMHM and a low mass threshold for galaxy formation. However, it is worth noting that cold tidal features, such as streams and shells, will have larger surface brightness than the spherically averaged profiles. Nonetheless, these features will only be seen at large projected distances when the dwarfs experience minor/intermediate mergers (i.e, for low mass threshold galaxy occupation models).

Note, here we only consider the CDM predictions, but there are also differences with the WDM models. The number of mergers are reduced in \textsc{coco-warm}, so, as for \textsc{coco-cold}, in most cases the dwarf stellar haloes are likely undetectable. However, while the signal will be reduced for Model B1 (typically 2 minor mergers in \textsc{coco-warm} relative to 6 in \textsc{coco-cold}), there is still a chance of detection when several dwarfs are stacked together. Note that also differences in halo concentration between WDM and CDM (we only consider $c_{200}=10$ here) could also lead to discernible differences in the observed profiles.
It is worth noting that there is an inevitable degeneracy between the dark matter particle and galaxy formation model: both WDM and a high mass threshold for galaxy formation can reduce minor mergers. However, there are still important differences between these scenarios, such as \textit{when} and how many merger events occur, which will produce different observational signatures. Of course, tangential lines of evidence, from, for example, properties of surviving dwarf galaxies will also help provide more quantitative assessments of any models.

\begin{figure*}
  \begin{minipage}{\linewidth}
        \centering
        \includegraphics[width=\textwidth,angle=0]{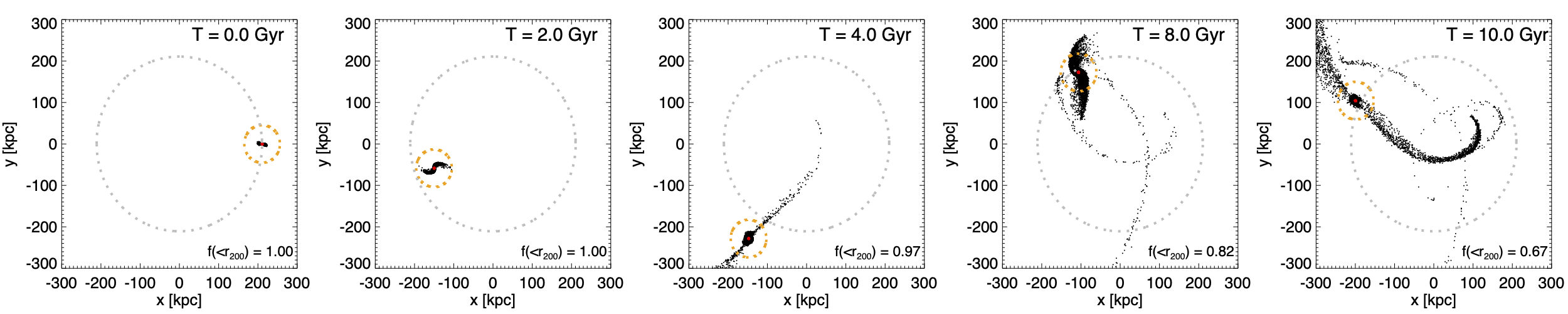}
    \end{minipage}
     \begin{minipage}{\linewidth}
       \centering
           \includegraphics[width=\textwidth,angle=0]{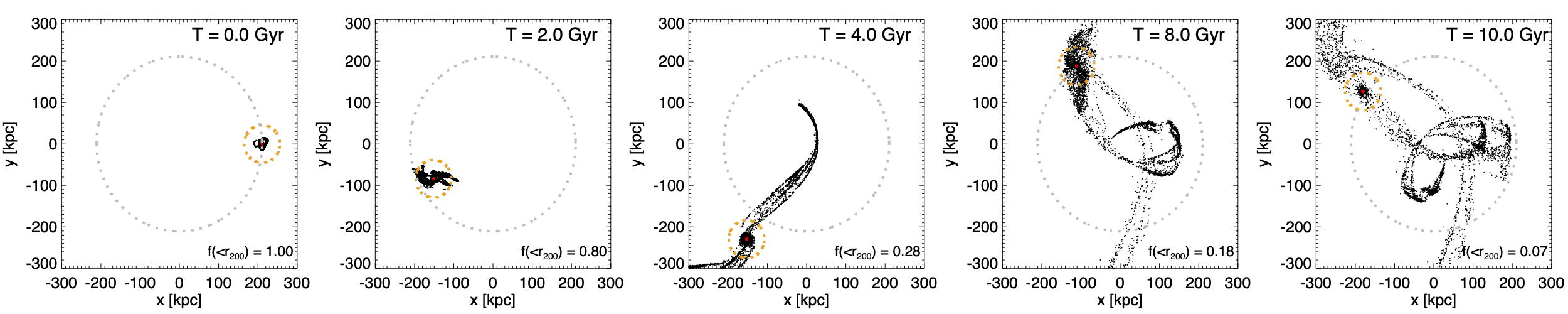}
       \end{minipage}
       \caption{The evolution of dwarf satellite stellar haloes in a Milky Way potential. Here, the dwarfs experience a 1:5 (top panel) or 1:10 (bottom panel) dark matter merger, and then fall into a Milky Way mass ($10^{12}M_\odot$) halo. The gray and orange dashed lines indicates the $r_{200}$ of the Milky Way, and dwarf satellite, respectively. In both cases, the central galaxy (red points) remains intact, but the dwarf stellar halo (black points) is stripped after $\sim 4$ Gyr. Interestingly, the path of the satellite stellar halo debris intersects with the central dwarf satellite galaxy. Thus, the existence of stellar streams overlapping in phase space with a known dwarf satellite galaxy could indicate that it was once a dwarf stellar halo.}
       \label{fig:mw_sats}
\end{figure*}

It is important to stress that there are several limitations to this exercise. First, we are only considering illustrative examples for each model, and there can be significant halo-to-halo variation. Second, our \textsc{Gadget} runs are isolated and do not allow for smooth dark matter mass growth. Thus, our analysis is based purely on a fixed $z=0$ halo, which is clearly not the case in reality. Clearly, the way forward is to employ a tagging scheme \textit{directly} onto a cosmological dark matter simulation suite.  This is beyond the scope of the current paper, but is something we aim to pursue in future work. Finally, we are also ignoring wet mergers, and the many effects this may have on the central galaxy, and, potentially, the stellar halo. Indeed, our dark matter tagging scheme may be missing important mechanisms that could ``kick out'' stars to the outer halo, from e.g. wet major mergers and/or feedback fluctuations. These can be explored in the future with  hydrodynamical simulations (see e.g. \citealt{tarumi21}).

\subsection{Streams associated with Milky Way dwarf satellites?}

\begin{figure*}
        \begin{minipage}{\linewidth}
        \centering
        \includegraphics[width=\textwidth,angle=0]{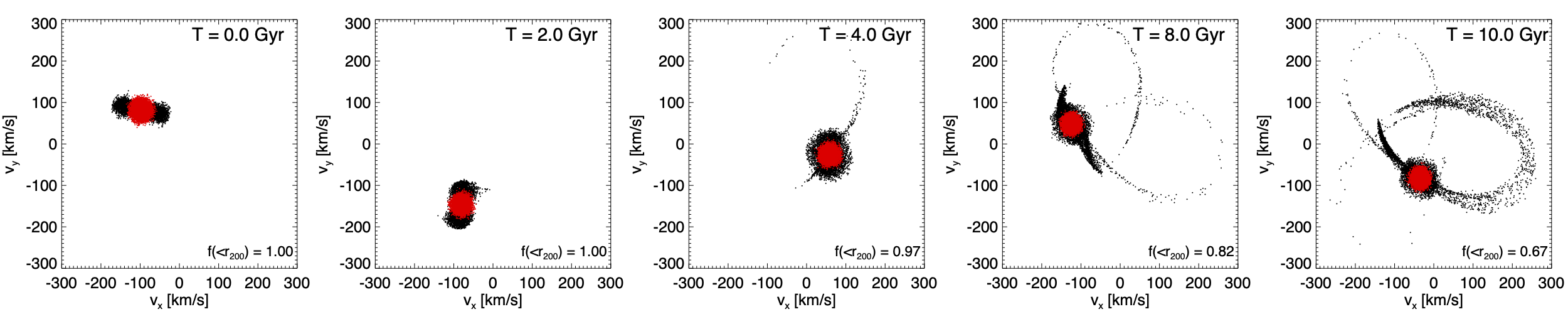}
    \end{minipage}
     \begin{minipage}{\linewidth}
       \centering
           \includegraphics[width=\textwidth,angle=0]{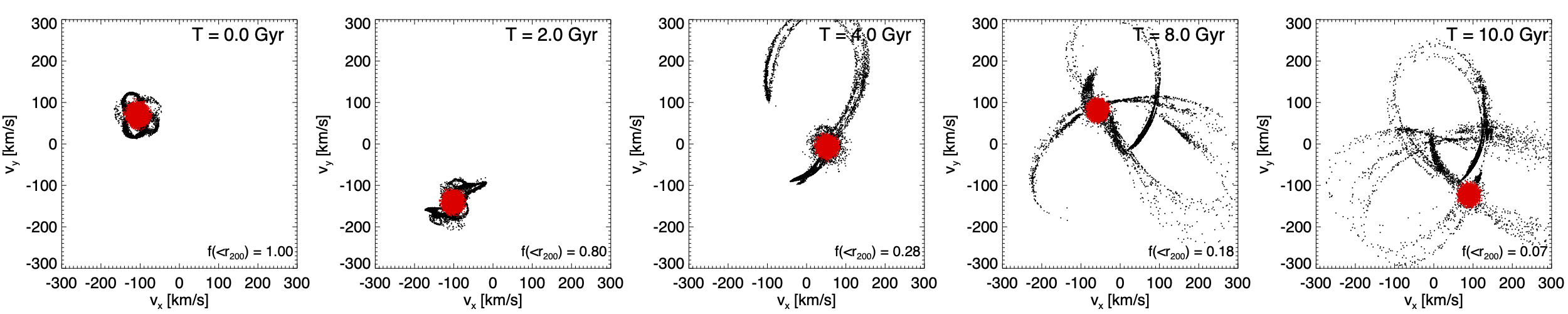}
       \end{minipage}
    \caption{As Fig. \ref{fig:mw_sats} but in velocity space.}
    \label{fig:mw_sats_vel}
\end{figure*}

Until now, we have considered dwarf merger events occurring in isolation. However, the most well-studied dwarf galaxies in our local vicinity are the satellites of the Milky Way. Indeed, there are several classical dwarfs that coincide with the halo mass scale that we have been studying ($10^{10}M_\odot$). Moreover, as we showed in Section \ref{sec:gals}, there can be important differences between the accreted stellar material of central and satellite dwarf galaxies with the same peak halo mass. This warrants the question: what happens to the stellar haloes of dwarf galaxies after falling into a MW mass halo? To address this question, we use two example mergers from our \textsc{Gadget} runs, and allow them to evolve inside a MW-mass potential. Thus, the initial conditions are determined by the (final) output of the dwarf-dwarf merger, and a $1\times 10^{12}M_\odot$ NFW halo (with $c_{200}=10$). As before, the initial orbital energy and angular momentum of the dwarf satellite system are chosen from the typical orbits in cosmological simulations \citep{jiang15}. For illustration, we show two examples, a 1:5 (dwarf-dwarf) dark matter merger, and a 1:10 (dwarf-dwarf) dark matter merger, which are shown in the bottom two panels of Fig. \ref{fig:examples}. In both cases, the central dwarf galaxy has a $10^{10}M_\odot$ halo, and the Milky Way host has a $10^{12}M_\odot$ halo.

Figures \ref{fig:mw_sats} and \ref{fig:mw_sats_vel} show how these systems evolve over time, both spatially and in velocity space. The red points are the (original) central dwarf galaxy, and the black points are the accreted dwarf stellar halo (no MW particles are shown).  Approximately, $3-4$ Gyr after infall the dwarf stellar halo begins to be tidally stripped. This is because this material is less bound than the central stellar component of the dwarf. After 10 Gyr a significant fraction of the dwarf stellar halo is stripped but the central dwarf galaxy remains in tact. Interestingly, the stripped dwarf stellar halo forms a stellar stream that intersects the central dwarf galaxy, both spatially and in velocity space. \textit{This suggests that the existence of a stellar stream overlapping in phase space with a known (surviving) dwarf satellite galaxy could indicate that it was once a dwarf stellar halo.} These streams also appear more structured in phase space than the streams that are usually formed from the bulk of the satellite itself, likely because they are formed of material that was not entirely phase-mixed to begin with. Importantly, this long-lived association is contrary to what would happen to ``satellites of satellites''. Here, the satellites of satellites very quickly (after a few Gyr) lose phase-space coherence with their original host \citep[see e.g][]{deason15}. In the case of ``satellite stellar haloes'', the coherence can remain in the form of stellar streams, even after several Gyr.

For comparison, we also perform the same exercise for a 1:2 (dwarf-dwarf) dark matter merger (shown in the top panel of Fig. \ref{fig:examples}). Here, the majority of the accreted dwarf stellar halo remains bound to the satellite, even $\sim 10$ Gyr after infall into the MW potential. The difference here is that the accreted material is more tightly bound to the dwarf, as it originates from a major merger. Thus, any extended stellar streams associated with known (surviving) dwarf satellites will likely be related to minor, rather than major, dwarf-dwarf mergers. And, as we have seen previously, the mere existence of (minor) dwarf-dwarf mergers significantly depends on the galaxy occupation model.

The potential existence of satellite stellar haloes in the MW halo is an exciting prospect, and in the current and upcoming era of \textit{Gaia} and wide-field spectroscopic surveys, such as DESI, WEAVE and 4MOST, this is a testable prediction. Indeed, the stellar halo of the (disrupting) Sagittarius has potentially already been uncovered by the H3 survey \citep[][but also see \citealt{penarrubia21}]{johnson20}. Interestingly, \cite{johnson20} identify the diffuse halo component by selecting Sagittarius stars in angular momentum space, in excellent agreement with the outputs of our toy models.  It is worth bearing in mind that there is a significant amount of parameter space to explore in our models (we only discuss a few examples here). As mentioned above, a fully-consistent tagging method applied to cosmological simulations will provide more quantitative predictions. Nonetheless, our current exploration suggests that the detection of these stripped satellite stellar haloes would strongly favour models with a low halo mass threshold for galaxy formation. 

\section{Conclusions}
\label{sec:conc}
In this work, we use a combination of cosmological N body simulations and empirical galaxy models to investigate the accreted stellar mass of dwarf-mass haloes ($M_{\rm halo} \sim 10^{10}M_\odot$). We also use simulations of isolated dwarf-dwarf mergers to infer how different types of (stellar) merger events build up dwarf galaxy stellar haloes. Our main conclusions are summarised as follows:

\begin{itemize}
    \item Dwarf-mass haloes ($M_{\rm halo} \sim 10^{10}M_\odot$ at $z=0$) in cold (warm) dark matter simulations typically experience $N=1.6 (0.6)$ major mergers, and $N=7 (2)$ minor mergers over their lifetime. Using empirical galaxy models we investigate how many of these mergers bring accreted stars into the halo. For models with high halo mass threshold for galaxy formation ($\sim 10^{9.3}M_\odot$ at $z=0$), minor mergers are largely suppressed, and the number of (stellar) merger events is reduced by a factor of $\sim 7-16$ relative to models with low halo mass threshold for galaxy formation ($\sim 10^{7.5}M_\odot$ at $z=0$).
    
    \item The galaxy models we consider predict drastically different stellar halo progenitors. Models with high halo mass threshold for galaxy formation, typically have a low number ($N \leq 1$) of major mergers at low/intermediate redshift ($z\sim 0-4$), while models with low mass threshold tend to have more minor mergers at relatively higher redshift ($z \sim 3-6$). In addition, models with a shallow SMHM slope typically have higher progenitor stellar masses than models with a steeper slope. Taken together, our results show that the contribution of accreted stars is strongly dependent on the galaxy model. The balance between minor and major mergers is particularly important as the distribution of accreted stars is largely determined by the satellite-central mass ratio.
    
    \item At dwarf mass scales, WDM models have a lower frequency of both major and minor mergers. However, differences between WDM and CDM are only seen in galaxy models with low halo mass thresholds for galaxy formation. If galaxies cannot occupy low mass haloes ($\lesssim 10^9M_\odot$) then there is little difference between the WDM and CDM predictions. Disentangling the affect of the dark matter particle and galaxy formation physics will likely require several lines of evidence. Our work shows that dwarf stellar haloes are an important, but relatively unexplored, probe.
    
    \item Satellite dwarfs at $z=0$ with the same peak halo mass ($\sim 10^{10}M_\odot$) have a different mass assembly history to centrals. They experience a similar number of total (major and minor) mergers, but these typically occur at higher redshift, and the satellite dwarfs commonly reach their peak mass earlier than centrals. As a consequence, satellites are able to undergo more stellar mergers at high redshift and tend to have richer stellar haloes than their central counterparts. However, these differences are modest for galaxy formation models with a low halo mass threshold, but are much more significant for models with a high halo mass threshold. Hence, the \textit{relative} differences between the accreted stellar populations of central and satellite dwarf systems could be a key discriminator of galaxy formation models.
    
    \item Using isolated models of dwarf-dwarf mergers we show that the fraction of accreted stars at large distances in the halo ($> 0.1r_{200}$) is dependent on the binding energy of the infalling satellite, and the amount of stellar material it contains. For example, major mergers can deposit a lot of stars, but owing to their high binding energy do not disperse many stars to very large distances in the halo. Conversely, minor mergers have much lower binding energy, and thus can deposit material at large distances, but they, by definition, bring in far fewer stars. In fact, ``intermediate'' mass mergers (with dark matter ratio $\sim$ 1:5) maximise the amount of stellar material deposited in the dwarfs' stellar halo, owing to this fine balance between binding energy and quantity of stars.
    
    \item In the galaxy models we consider, major mergers are not uncommon ($N \sim 0.4-1.5$ per halo). This is in agreement with observations, as there are several lines of evidence suggesting major mergers have occurred in some known dwarf galaxies \citep[e.g.][]{amorisco12, amorisco14}. While our isolated simulations suggest that these major mergers deposit little material in the stellar halo (or at least beyond $0.1r_{200}$), we are not taking into account gas-rich mergers, which could enhance the amount of material at large distances. Nonetheless, we conclude that it is minor mergers which likely hold the strongest clues for dwarf galaxy models. Indeed, the very existence of minor merger features could rule out galaxy formation models with a high halo mass threshold.

    \item We consider the observability of dwarf galaxy stellar haloes, and find that the surface brightness levels are likely well below feasible detection limits. However, there is some hope if several dwarf galaxies can be stacked together. Nonetheless, only models with a low halo mass threshold for galaxy formation, and a relatively shallow SMHM relation will likely produce observable surface brightness features. The absence of such features may rule these models out.
    
    \item Finally, we consider what happens to dwarf stellar haloes after they infall into a Milky Way mass halo. Interestingly, unlike satellites-of-satellites that quickly disperse in phase space, dwarf stellar haloes can form stellar streams, which overlap in phase space with the central (surviving) dwarf galaxy. Thus,  observational probes of stellar streams linked to known dwarf galaxy satellites in the Milky Way could be the most feasible method to detect dwarf galaxy stellar haloes.

\end{itemize}

We have found that the properties of dwarf galaxy stellar haloes are very sensitive to low-mass galaxy formation models and the type of  dark matter particle. Indeed, the mere existence of these features will likely rule out a significant amount of model parameter space. Observational probes are challenging because of the incredibly low surface brightness of these diffuse haloes. However, there is hope that with stacking and/or detection of cold features these stellar haloes can be revealed. Moreover, resolved stellar populations, such as in our Local Group, hold a lot of promise. For example, the stellar streams of ripped apart dwarf stellar haloes can remain coherent in phase space with the central galaxy for sometime; thus known Milky Way satellites may be the best testing ground. Of course, we also have more work to do on the theoretical side, as we have taken a relatively simple approach in this work. Nonetheless, the search for dwarf stellar haloes is on, and they may hold the strongest clues yet as to the nature of galaxy formation and dark matter on the lowest mass scales.

\section*{Acknowledgements}
 AD thanks the staff at the Durham University Day Nursery who play a key role in enabling research like this to happen.
 
\noindent
AD is supported by a Royal Society University Research Fellowship. AD acknowledges support from the Leverhulme Trust and the Science and Technology Facilities Council (STFC) [grant numbers ST/P000541/1, ST/T000244/1]. SB and AF are supported by the UK Research and Innovation (UKRI) Future Leaders Fellowships (grant numbers MR/V023381/1, MR/T042362/1). NA is supported by an STFC/UKRI Ernest Rutherford
Fellowship, Project Reference: ST/S004998/1.
WH acknowledges support of the National Science Center, Poland under projects no. UMO-2018/31/G/ST9/03388 and UMO-2020/39/B/ST9/03494. CSF acknowledges European Research
Council (ERC) Advanced Investigator grant DMIDAS (GA 786910). 
\noindent
This work used the DiRAC@Durham facility managed by the Institute for Computational Cosmology on behalf of the STFC DiRAC HPC Facility (\url{www.dirac.ac.uk}). The equipment was funded by BEIS capital funding via STFC capital grants ST/K00042X/1, ST/P002293/1, ST/R002371/1 and ST/S002502/1, Durham University and STFC operations grant ST/R000832/1. DiRAC is part of the National e-Infrastructure.

\section*{Data Availability}
The data presented in the figures are available upon request from the corresponding author.

\bibliographystyle{mnras}
\bibliography{mybib}

\bsp
\label{lastpage}
\end{document}